\documentclass[aoas]{imsart}

\RequirePackage{amsthm,amsmath,amsfonts,amssymb}
\RequirePackage[authoryear]{natbib}
\RequirePackage[colorlinks,citecolor=blue,urlcolor=blue]{hyperref}
\RequirePackage{graphicx}

\startlocaldefs
\usepackage{threeparttable}
\usepackage{graphicx}%
\usepackage{multirow}%
\usepackage{amsmath,amssymb,amsfonts}

\usepackage{amsthm}%
\usepackage{mathrsfs}%
\usepackage[title]{appendix}%
\usepackage{xcolor}%
\usepackage{textcomp}%
\usepackage{manyfoot}%
\usepackage{booktabs}%
\usepackage{algorithm}%
\usepackage{listings}%
\usepackage{graphicx}
\usepackage{booktabs}

\usepackage{adjustbox}
\usepackage{amsmath,amsfonts,amsthm,bm} 

\usepackage{graphicx} 
\usepackage{amsmath,amsfonts}
\usepackage{algorithmic}
\usepackage{algorithm}
\usepackage{array}
\usepackage{textcomp}
\usepackage{stfloats}
\usepackage{mathabx}
\usepackage{url}
\usepackage{verbatim}
\usepackage{graphicx}
\usepackage{multirow}
\usepackage{multicol}
\usepackage{xcolor}
\usepackage{microtype}
\usepackage{longtable}
\usepackage{tabto}
\usepackage{graphicx}
\usepackage{tabularx}
\usepackage{amsmath}
\usepackage{float}
\usepackage{subcaption}
\usepackage{caption}
\usepackage{adjustbox}
\usepackage{tikz}
\usepackage[normalem]{ulem}
\useunder{\uline}{\ul}{}

\numberwithin{equation}{section}
\theoremstyle{plain}

\usepackage{xcolor}

\definecolor{royalpurple}{rgb}{0.47, 0.32, 0.66}
\definecolor{Raspberry}{RGB}{227, 11, 92}
\definecolor{changeblue}{RGB}{65,135,195}
\definecolor{revisionorange}{RGB}{204,102,0} 

\usepackage{color}

\newcommand{\ncolor}[1]{\textcolor{teal}{#1}}  

\newcommand{\rcolor}[1]{\textcolor{red}{#1}}

\usepackage{xcolor}
\usepackage[colorinlistoftodos]{todonotes}
\newcommand{\bea}{\begin{eqnarray}}
\newcommand{\ena}{\end{eqnarray}}
\newcommand{\beas}{\begin{eqnarray*}}
\newcommand{\enas}{\end{eqnarray*}}
\newcommand{\beq}{\begin{equation}}
\newcommand{\enq}{\end{equation}}
\def\qed{\hfill \mbox{\rule{0.5em}{0.5em}}}

\newcommand{\ignore}[1]{}

\newcommand{\qmq}[1]{\quad\mbox{#1}\quad}

\DeclareMathOperator*{\argmin}{arg\,min}

\makeatletter
\newsavebox\myboxA
\newsavebox\myboxB
\newlength\mylenA
\newcommand*\xoverline[2][0.70]{%
    \sbox{\myboxA}{$\m@th#2$}%
    \setbox\myboxB\null
    \ht\myboxB=\ht\myboxA%
    \dp\myboxB=\dp\myboxA%
    \wd\myboxB=#1\wd\myboxA
    \sbox\myboxB{$\m@th\overline{\copy\myboxB}$}
    \setlength\mylenA{\the\wd\myboxA}
    \addtolength\mylenA{-\the\wd\myboxB}%
    \ifdim\wd\myboxB<\wd\myboxA%
       \rlap{\hskip 0.5\mylenA\usebox\myboxB}{\usebox\myboxA}%
    \else
        \hskip -0.5\mylenA\rlap{\usebox\myboxA}{\hskip 0.5\mylenA\usebox\myboxB}%
    \fi}
\makeatother

\theoremstyle{plain}

\newtheorem{theorem}{Theorem}[section]

\newtheorem{proposition}[theorem]{Proposition}
\theoremstyle{definition}

\endlocaldefs

\usepackage{neutralpreprint}

\begin{document}

\begin{frontmatter}
\title{$K$-Midranks: A robust clustering algorithm for heavy-tailed data with applications to spectroscopy and finance}
\runtitle{$K$-Midranks: A robust clustering algorithm}

\begin{aug}
\author[A]{\fnms{Larry}~\snm{Goldstein}\ead[label=e1]{larry@usc.edu}}
\author[B]{\fnms{Nathakhun}~\snm{Wiroonsri}
\thanks{\textbf{Corresponding author.}}\ead[label=e2]{nathakhun.wir@kmutt.edu}\orcid{0000-0003-2167-9641}}
\address[A]{Department of Mathematics,
University of Southern California, USA \printead[presep={ ,\ }]{e1}}

\address[B]{Department of Mathematics, King Mongkut's University of Technology Thonburi, Thailand \printead[presep={,\ }]{e2}}
\end{aug}

\begin{abstract}
Clustering data with extreme observations and heavy tails can be challenging for methods that rely on distances between raw observations or specific distributional assumptions. We introduce $K$-midranks, a clustering method that replaces observations with coordinatewise ranks and partitions the transformed data using a procedure analogous to $K$-means. This transformation reduces the influence of extreme values while preserving the ordering of observations within each variable. We evaluate $K$-midranks against $K$-means, $K$-medoids, and mixture models based on Gaussian, Student's t, Cauchy, and generalized hyperbolic distributions through simulation studies, and provide some theoretical properties. $K$-midranks generally outperforms the competing methods on Cauchy-shifted data. In contrast, $K$-means and model-based methods perform better on Gaussian data and Student's t data with three degrees of freedom. We demonstrate the method's practical use in grouping samples based on Raman and LIBS spectra and clustering S\&P 500 stocks according to their returns. $K$-midranks achieves competitive performance with model-based methods in the spectroscopy applications and the best performance among the methods considered in the stock-return application. These findings suggest that $K$-midranks provides a useful alternative for clustering data with Cauchy-like heavy tails. An R package implementing the proposed method is available.

\end{abstract}

\begin{keyword}
\kwd{First keyword}
\kwd{second keyword}
\end{keyword}

\end{frontmatter}

\section{Introduction}

\label{sec:introduction}


Cluster analysis is a fundamental process in statistics and machine learning in which data are grouped into subsets, known as clusters, according to their similarity (see \citet{statbook2021} for a review). The primary goal is to uncover hidden structure in data not having group labels. Cluster analysis is widely applied across fields including business, medicine, and the environmental and social sciences. For instance, it is used to identify groups of customers with similar purchasing behavior, to group cancer cells by biochemical characteristics, and analyze wildfire risk areas using satellite data (see \citet{exbio,exfire} for more examples). The well known $K$-means clustering \citep{kmean} algorithm is perhaps the most notable, though the collection also includes,  $K$-medoids \citep{Kaufman90}, hierarchical clustering \citep{hierarchy}, 
EM clustering \citep{em}, DBSCAN \citep{dbscan}, and model based clustering \citep{banfield1993model}. 



While many algorithms may be reliable when applied to well behaved data sets, they may fail when the observations are governed by distributions having heavy tails. 
Such distributions commonly arise in many real-world applications, including financial returns, spectroscopic measurements (e.g., Raman, LIBS, and mass spectrometry), insurance claim sizes, telecommunication and network traffic, environmental extremes (e.g., rainfall and flood events), and biomedical data (see \cite{cont2001empirical, embrechts2013modelling, coles2001introduction}). In these applications, the presence of extreme observations is an inherent characteristic of the data rather than a result of measurement errors.

Heavy-tailed data present significant challenges for conventional clustering algorithms such as $K$-means and $K$-medoids, which are sensitive to extreme observations \citep{jain2010data, garcia2010review}. Since these algorithms optimize distance-based objective functions, a small number of extreme observations can substantially distort the estimated cluster centers, leading to inaccurate cluster assignments, especially when clusters overlap. Consequently, specialized clustering methods are required to analyze datasets exhibiting heavy-tailed characteristics effectively. The most popular option is model-based clustering \citep{andrews2012model, mcnicholas2016mixture, zhang2023model}. However, its performance relies heavily on the choice of the underlying probability distribution, which may negatively affect the convergence of the procedure's parameter estimates, particularly for complex models or high-dimensional data.

In this work, we introduce $K$-midranks, a novel clustering method based on coordinatewise ranks. 
Replacing continuous variables with coordinatewise ranks reduces the influence of outliers and extreme values, and does not require assumptions about the existence of marginal moments.

Transformations to ranks have a storied history in non-parametric statistics, including the well known rank correlation of Spearman \citep{Spearman1904} and the
Wilcoxon rank-sum/Mann--Whitney test
\citep{Wilcoxon1945,MannWhitney1947}.
In all such procedures, the gain in robustness and stability in the presence of outliers is made at the cost of discarding the sizes of
the gaps between observations, which may contain useful information.

The connection between clustering and rank-based techniques has been studied previously. \cite{cape2024robust} proposed an eigenvector-based clustering method that operates on a matrix of nonparametric normalized rank statistics derived entrywise from the original data matrix. Unlike our setting, the raw data in their work take the form of an $n \times n$ weighted adjacency matrix, where each entry represents a random relationship between objects $i$ and $j$. A symmetric rank transformation is first applied to the original matrix, after which clustering is performed using the truncated eigendecomposition of the resulting rank matrix. \cite{ballante2024fully} proposed a hierarchical clustering algorithm based on the rank-based similarity measure introduced by \cite{ballante2024non}, which extends the approach of \cite{glielmo2022ranking}. In their framework, the rank $r_{ij}$ is defined as the rank of object $j$ from the perspective of object $i$. These definitions differ fundamentally from the coordinatewise rank employed here in our proposed method, which is introduced below.

We assess the performance of the proposed $K$-midranks algorithm by first evaluating it on simulated datasets with known class labels and comparing its performance with several existing clustering methods, including $K$-means, $K$-medoids, and model-based clustering based on Gaussian, Student's t, Cauchy, and generalized hyperbolic distributions. We then apply $K$-midranks to the real-world datasets mentioned above.


The remainder of this work is organized as follows.
Section~\ref{sec:background} gives the scientific and data context.
The proposed methodology is presented in Section~\ref{sec:main}, followed by
the main mathematical and statistical properties in Section~\ref{sec:properties}.
Section~\ref{sec:exp} presents the simulation and comparative validation study,
and Section~\ref{sec:app} gives the real-data analyses.
Discussion, limitations, and concluding remarks appear in
Section~\ref{sec:conclusion}. A supplement to this article provides the proofs of the claims made regarding the procedure. 

\section{Scientific and data context}\label{sec:background}

\subsection{Heavy-tailed distributions}

A heavy-tailed distribution is a probability distribution whose tail probabilities decay more slowly than those of an exponential distribution. Consequently, extreme observations occur with a substantially higher probability than under light-tailed distributions, such as the normal distribution. 
Many common heavy-tailed distributions lack a finite first absolute moment, including the Cauchy distribution, the Pareto distribution with $0<\alpha<1$, and Student's t distribution with $0\le \nu \le 1$.

\subsection{Scientific application settings}
We introduce the real-world settings considered in the application section of this study. Specifically, we consider Laser-Induced Breakdown Spectroscopy (LIBS), Raman spectroscopy, and Standard \& Poor's 500 (S\&P 500) stock returns.

Spectroscopy is one of the principal tools by which scientists infer the
physical and chemical properties of matter from its interaction with
electromagnetic radiation. Different atoms and molecules absorb, emit, or
scatter radiation at characteristic wavelengths, so that its measured spectrum
can contain information about many of its physical properties, and principally its chemical composition. Spectroscopy is
also useful when direct examination of an object is difficult or impossible. In astronomy, for example, spectra
are used to determine the chemical composition and motion of stars and
galaxies and to identify molecules in the atmospheres of distant planets.

Raman and LIBS spectroscopy provide complementary information about
materials. Raman spectroscopy \citep{flanagan2025open} probes molecular
structure through small frequency shifts in laser light scattered by a
sample, while LIBS \citep{kepevs2020benchmark} creates a short-lived plasma
and analyzes the light emitted as it cools, providing information principally
about elemental composition. The physical mechanisms of the two techniques
are described in more detail in Sections~\ref{subsec:raman} and
\ref{subsec:libs}.

The growing use of spectroscopy also creates a statistical problem: a single
experiment may produce hundreds or thousands of spectral measurements, each
recorded at thousands of wavelengths. Clustering provides a way to explore
such data without prespecifying the identities of the samples. If spectra
naturally separate into groups, they may correspond to a shared property such as composition, or forms. Clustering can therefore be used not only to
compress a large collection of spectra into an interpretable structure, but
also to reveal unexpected similarities that could merit further
investigation.

Because $K$-midranks is intended primarily for settings in which
extreme observations can make standard methods unreliable, it is important to examine whether the numerical
features in our applications exhibit heavy tails. We
investigate this empirically using normal Q--Q plots. We first compare the Q--Q behavior of the Gaussian,
Student's $t$, and Cauchy-based simulated datasets, and then compare these
patterns with those observed in the Raman, LIBS, and S\&P 500 data; see
Figures~\ref{fig:qq1} and~\ref{fig:qq2}. These diagnostics provide visual evidence
about the presence of extreme tail behavior, but are not intended by
themselves to establish that a population moment is infinite.

Next, the well known S\&P 500 stock market index consists of 500 leading publicly traded companies in the United States and is widely regarded as a proxy for the overall U.S. equity market. Return distributions are well known to exhibit heavy tails \citep{borak2011models}, making them challenging for conventional statistical methods that are predicated on Gaussian tails. 

Although S\&P 500 companies are commonly categorized into sectors such as information technology, healthcare, and financials, these classifications are based on primary business activities rather than market behavior. Stocks from different sectors may exhibit similar return patterns because they are exposed to common economic factors, while companies within the same sector may behave differently due to differences in size, growth prospects, or financial structure. Consequently, clustering can reveal latent market structure that can identify groups of stocks with similar risk and return profiles.

\section{Proposed methodology}\label{sec:main}

\subsection{$K$-midranks and other clustering methods based on representatives} 
\label{subsec:representative}

Many clustering methods determine a partition 
${\bm C}=\{C_1,\ldots,C_K\}$ of $[n]$ into clusters, with each set in the collection non-empty and all pairs having null intersection, by minimizing 
an objective function $f$ depending on a choice of dissimilarity measure $d$ and of center, or representative $\boldsymbol{\nu}_C$, for the non-empty cluster $C$. 

We first introduce $K$-midranks, followed by the methods we use for performance comparison. The key difference between $K$-midranks and the other methods considered is that it replaces the given data by coordinatewise ranks before computing distances and cluster representatives. This transformation removes marginal units and is independent of the sizes of marginal gaps. This makes $K$-midranks more robust against extreme values and outliers, as the median is with respect to the mean. However, as with other procedures that protect against outliers, there is typically an efficiency price to pay when the data are well behaved, such as when their tails are sub-Gaussian.

In the following, we will use 
$$
d_q(\mathbf y,\mathbf z)
 =\left(\sum_{j=1}^p |y_j-z_j|^q\right)^{1/q}, \quad q =1,2
$$
to denote the $L^1$ and Euclidean distance, respectively. 

\subsubsection{$K$-midranks} 
Here we first transform the data matrix $\mathbf{x} = [x_{ij}]_{i \in [n], j\in [p]}$
using ranks, define the column-wise rank matrix ${\bm R} = [R_{ij}]_{i\in[n], j\in[p]}$to have entries 
\bea \label{rkdef}
R_{ij} = 1+\sum_{k=1}^n\mathbf{1}(x_{kj}>x_{ij}) \qmq{and write}{\bm R}_i = (R_{i1},R_{i2},\ldots,R_{ip}).
\ena
Notice that the largest value is given rank 1, and equal observations are given the same rank.  

For $q\in\{1,2\}$, the objective function and cluster representatives
are respectively given by
\begin{align}\label{eq:kmr-lqloss}
f(\boldsymbol C)
=
\sum_{k=1}^K\sum_{i\in C_k}
d_q^q({\bm R}_i,\boldsymbol{\nu}_k),
\quad \mbox{and} \quad
\boldsymbol{\nu}_k
\in
\argmin_{\mathbf z\in\mathbb R^p}
\sum_{i\in C_k}d_q^q({\bm R}_i,\mathbf z),
\qquad k=1,\ldots,K,
\end{align}
where the minimization is 
over partitions $\boldsymbol{C}=\{C_1,\ldots,C_K\}$ of $[n]$ where each set is non-empty and all pairs have null intersection.
When  $q=1$, any coordinatewise median $\boldsymbol{\nu}_{C}$ of the rank vectors in $C$ is a
minimizer in \eqref{eq:kmr-lqloss}; for $q=2$, the
unique minimizer is the coordinatewise mean. The algorithm we use in practice to minimize the objective function in \eqref{eq:kmr-lqloss} is given at the end of this section.

\subsubsection{$K$-means}

The objective function $f$ and choice of cluster representative of the $K$-means method \citep{kmean} are given by
\begin{equation}\label{def.fKmeans}
 f({\bm C}) =   \sum_{k=1}^K\sum_{i \in C_k} d_2^2(\mathbf{x}_i,\boldsymbol{\nu}_k) \qmq{and} \boldsymbol{\nu}_k = \frac{1}{|C_k|}\sum_{i \in C_k} \mathbf{x}_i.
\end{equation}
The squared-Euclidean version of $K$-midranks uses the same
within-cluster sum-of-squares criterion, but applies it to the
coordinatewise rank vectors rather than to the original observations.
\subsubsection{$K$-medians}
The objective function $f$ and choice of cluster representative of $K$-medians \citep{jain1988algorithms} are given by 
\begin{equation} 
 f({\bm C}) =   \sum_{k=1}^K\sum_{i \in C_k} d_1(\mathbf{x}_i,\boldsymbol{\nu}_k) \qmq{and} \boldsymbol{\nu}_k = ({\rm median}(x_{ij},i \in C_k))_{1 \le j \le p}, 
\end{equation}
where $x_{ij}$ denotes the $j$th coordinate of observation $x_i$. 
Hence this method is similar to $K$-means in \eqref{def.fKmeans},
but replaces the squared Euclidean loss $d_2^2$ by the $L^1$ loss
$d_1$, and the mean representative by the coordinatewise median.

Since the median and the size of an absolute deviation are less sensitive to extreme observations than the mean and squared deviations, $K$-medians is generally more robust to outliers than $K$-means.

The $L^1$ version of $K$-midranks is $K$-medians applied to the
coordinatewise rank vectors. Thus both use coordinatewise medians as
representatives, but ordinary $K$-medians retains the original
numerical magnitudes, whereas $K$-midranks replaces them by their
relative order.

\subsubsection{$K$-medoids} 
$K$-medoids also partitions the data into $K$ groups, but it represents each cluster by an observed data point, called its medoid.  For a given dissimilarity $d$, the objective function and the choice of center are respectively given by
\[
 f(\bm C) = \sum_{k=1}^K\sum_{i\in C_k}d(\mathbf{x}_i,\boldsymbol{\nu}_k) \qmq{and} \boldsymbol{\nu}_k \in
\argmin_{\mathbf z\in\{\mathbf x_h:h\in C_k\}}
\sum_{i\in C_k}d(\mathbf x_i,\mathbf z).
\]

 One common choice for the measure $d$ is the $L^1$ distance. We note that here the cluster representative is an observed data point. Consequently $K$-medoids does not require computing, say, averages or medians. For a fixed cluster and data representation, restricting the
representative to an observed point can give a larger minimum
$L^1$ loss than allowing an arbitrary representative, as in
$K$-medians and $K$-midranks. Although $K$-medoids is usually less
sensitive than $K$-means to isolated extreme observations, large
distances between the original observations can still influence its
assignments and objective. In contrast, $K$-midranks computes
dissimilarities after replacing the observations by coordinatewise
ranks.

\subsection{Model-based clustering} 
Model-based clustering \citep{banfield1993model} assumes that the observations arise from a finite mixture of probability distributions, whose parameters may differ across components to capture differences in location, spread, and shape (see \cite{mcnicholas2016mixture} for a review). The mixing proportions and the parameters of the component distributions are estimated from the data. The fitted model is then used to calculate the posterior probability that each observation belongs to each component. These probabilities provide soft cluster memberships, while a hard clustering is obtained by assigning each observation to the component with the largest posterior probability. 

Examples include Gaussian, Student's $t$, and generalized hyperbolic mixture models, which differ in the distributions used to model the latent groups. Gaussian components describe ellipsoidal groups, while Student's $t$ components accommodate heavier tails. Generalized hyperbolic components allow additional flexibility in tail behavior and asymmetry.

In comparison with ours, Model-based methods retain the original numerical measurements and posit an explicit probability model. On the positive side, they can accommodate overlapping clusters, quantify posterior membership uncertainty, and estimate cluster-specific location and shape parameters. Their performance, however, depends on whether the particular distributional family selected by the user is actually in force, and how reliably its parameters can be estimated. These methods may employ 
heavy-tailed distributions such as the Student's t, the generalized hyperbolic, and Cauchy distributions to accommodate 
extreme observations. In contrast, the use of $K$-midranks does not require any parametric model for the data to be adopted, and achieves robustness by transforming the data via ranks, which depend only on the ordering of observations which are are independent of the sizes of marginal gaps.

To run $K$-midranks in practice, we use Algorithm \ref{alg:kmid}, modeled after that for $K$-means. In particular, the algorithm is essentially the usual $K$-means algorithm applied to data whose coordinate values have been replaced by ranks, and modified to have the two loss functions we consider; see Proposition \ref{prop:kmr-current-centers}. We include the algorithm description not only for clarity and completeness, but also so that we can compute its complexity in Proposition \ref{prop:kmr-current-algorithm} below.

\begin{algorithm}[H]
\caption{$K$-midranks Clustering}
\label{alg:kmid}
\begin{algorithmic}[1]
\REQUIRE Data matrix $X=(x_{ij})\in\mathbb R^{n\times p}$, positive
integers $n,p$, an integer $1\le K\le n$, and a loss $d$ as above.
\ENSURE Assignments $c_1,\ldots,c_n$ to $K$ nonempty clusters and
representatives $\mathbf v_1,\ldots,\mathbf v_K$.
\STATE Construct $\bm R_i=(R_{i1},\ldots,R_{ip})$ using descending
columnwise ranks
\[
R_{ij}=1+\sum_{h=1}^n\mathbf 1\{x_{hj}>x_{ij}\}.
\]
\STATE Initialize $c_1,\ldots,c_n$ so that every cluster is nonempty.
\REPEAT
    \STATE Save $\mathbf c^{\mathrm{old}}\leftarrow(c_1,\ldots,c_n)$.
    \FOR{$k=1,\ldots,K$}
        \STATE Choose a minimizing representative
        \[
        \mathbf v_k\in\operatorname*{arg\,min}_{\mathbf z\in\mathbb R^p}
        \sum_{i\in C_k} d(\bm R_i,\mathbf z).
        \]
        For $L^1$ loss, take the coordinatewise median, choosing the
        midpoint of the two central ordered values whenever $|C_k|$
        is even. For squared $L^2$ loss, take the coordinatewise mean.
    \ENDFOR
    \STATE Keep the representatives fixed throughout the next
    assignment step.
    \FOR{$i=1,\ldots,n$}
        \STATE Let $M_i$ be the set of minimizing cluster labels:
        \[
        M_i=\left\{k\in\{1,\ldots,K\}:
        d(\bm R_i,\mathbf v_k)=
        \min_{1\le h\le K}d(\bm R_i,\mathbf v_h)\right\}.
        \]
        If $c_i\in M_i$, retain the current assignment; otherwise
        set $c_i\leftarrow\min M_i$.
    \ENDFOR
    \WHILE{some cluster is empty}
        \STATE Let $k$ be the smallest label of an empty cluster.
        \STATE Choose the smallest index $i$ whose current cluster
        contains at least two observations.
        \STATE Set $c_i\leftarrow k$ and $\mathbf v_k\leftarrow\bm R_i$,
        (thus updating the cluster memberships and sizes.)
    \ENDWHILE
\UNTIL{$(c_1,\ldots,c_n)=\mathbf c^{\mathrm{old}}$} 
\end{algorithmic}
\end{algorithm}

\section{Mathematical and statistical properties}\label{sec:properties}

\noindent\textit{Proofs of the propositions stated in this section, together
with additional mathematical results, are given in the Supplementary
Material.}

For convenience, define the normalized rank vectors by
\begin{equation}\label{eq:kmr-normalized-ranks-current}
\mathbf u_i=(u_{i1},\ldots,u_{ip}),
\qquad
u_{ij}=\frac{R_{ij}}{n+1},
\qquad i\in[n],\quad j\in[p].
\end{equation}
For each fixed partition and $q\in\{1,2\}$, replacing the rank
vectors by their normalized versions multiplies the objective
in \eqref{eq:kmr-lqloss}, minimized over the representatives,
by $(n+1)^{-q}$. Consequently, normalization leaves the set of
minimizing partitions unchanged. The corresponding representatives
are divided by $n+1$. The normalization is useful because every
$\mathbf u_i$ lies in $(0,1)^p$.

\begin{proposition}[Form of the cluster representatives]
\label{prop:kmr-current-centers}
Let $C\subseteq[n]$ be nonempty. For $q\in\{1,2\}$, consider
the normalized cluster representatives defined by
\begin{equation}\label{eq:kmr-cluster-representative}
\mathbf v_C\in
\argmin_{\mathbf z\in\mathbb R^p}
\sum_{i\in C}d_q^q(\mathbf u_i,\mathbf z).
\end{equation}
\begin{enumerate}
\item
For $q=1$, the minimization separates by coordinates, and the
minimizers are exactly the coordinatewise medians. More precisely,
if
\[
u_{(1:j)}^{C}\le\cdots\le u_{(|C|:j)}^{C}
\]
are the ordered values of coordinate $j$ among the normalized
rank vectors indexed by $C$, then $\mathbf v_C$ is a minimizing
representative if and only if
\[
v_{C,j}\in
\left[
u_{(\lceil |C|/2\rceil:j)}^{C},
u_{(\lfloor |C|/2\rfloor+1:j)}^{C}
\right],
\qquad j\in[p].
\]
We take the midpoint of this interval in each coordinate.
Thus the $L^1$ version of $K$-midranks is $K$-medians applied
to the normalized rank vectors.

\item
For $q=2$, the unique minimizing representative is the mean
\[
\mathbf v_C
=
\frac{1}{|C|}\sum_{i\in C}\mathbf u_i.
\]
Thus the squared-Euclidean version of $K$-midranks is $K$-means
applied to the normalized rank vectors.
\end{enumerate}
\end{proposition}

\begin{proposition}[Monotone invariance]
\label{prop:kmr-current-invariance}
For each coordinate $j=1,\ldots,p$, let $\phi_j$ be a strictly increasing function. Then the replacement of 
$x_{ij}$ by $\phi_j(x_{ij})$ leaves the rank matrix in \eqref{rkdef} unchanged. In particular, for the losses
\eqref{eq:kmr-lqloss} with $q \in \{1,2\}$ the objective function and
the set of globally minimizing partitions are unchanged.

If every coordinate is tie-free, the objective and the set of globally
minimizing partitions are likewise unchanged when each $\phi_j$ is 
strictly decreasing. For a decreasing $\phi_j$ one
then has
\begin{equation}\label{eq:kmr-current-reflection}
 R_{ij}'=n+1-R_{ij},
 \qquad
 u_{ij}'=1-u_{ij},
\end{equation}
and coordinatewise reflections preserve both $L^1$ and squared-Euclidean
dissimilarities.
\end{proposition}

It is also clear that the method is
invariant under permutations of the coordinates.

\subsection{Balanced rowwise Rand distance robustness}
\label{subsec:balanced-row-half-rand}

We now demonstrate that, for the balanced two-cluster benchmark
defined below, $K$-midranks with squared-Euclidean loss and global
minimization over balanced partitions has a ``breakdown point''
comparable to that of the usual median. For two partitions $\mathcal P$ and
$\mathcal Q$ of $[n]$, define their Rand distance by
\begin{equation}
d_{\mathrm R}(\mathcal P,\mathcal Q)
=
\frac1{\binom n2}
\sum_{1\le i<\ell\le n}
\left|
\mathbf 1\{i\sim_{\mathcal P}\ell\}
-
\mathbf 1\{i\sim_{\mathcal Q}\ell\}
\right|.
\label{eq:balanced-half-rand-distance}
\end{equation}
Thus $d_{\mathrm R}(\mathcal P,\mathcal Q)\ge 1/2$ means that the
two partitions disagree on at least half of all pairwise co-membership
decisions.

Let $n\ge4$ be even and write $n=2m$. Let
$\mathcal P=\{P_A,P_B\}$, where $P_A$ and $P_B$ are disjoint and
both have size $m$. Consider the benchmark matrix $X=X^{(n,p)}$ with
entries
\begin{equation}
x_{ij}
=
\begin{cases}
1,&i\in P_A,\\
0,&i\in P_B,
\end{cases}
\qquad i\in[n],\quad j\in[p].
\label{eq:balanced-half-rand-benchmark}
\end{equation}
Every coordinate perfectly separates the two groups. Consequently,
$\mathcal P$ is the unique two-block global minimizer, up to relabeling.

For a data matrix $Y \in \mathbb{R}^{n \times p}$, let $\mathbf u_i(Y)$ denote the normalized rank vector
of its $i$th row, computed as in \eqref{eq:kmr-normalized-ranks-current}.
For a two-block partition $\mathcal R=\{R_1,R_2\}$, define
\begin{equation}
F_Y(\mathcal R)
=
\sum_{h=1}^2\sum_{i\in R_h}
\left\|
\mathbf u_i(Y)-\overline{\mathbf u}_{R_h}(Y)
\right\|_2^2,
\qquad
\overline{\mathbf u}_{R_h}(Y)
=
\frac1{|R_h|}
\sum_{i\in R_h}\mathbf u_i(Y).
\label{eq:balanced-half-rand-objective}
\end{equation}
Thus $F_Y$ is the squared-Euclidean $K$-midranks objective, with the
normalized rank vectors computed from $Y$.

Let $\Pi_{n,2}^{\mathrm{bal}}$ be the collection of partitions of
$[n]$ into two blocks of size $m$, and define the set of globally
minimizing balanced partitions by
\[
\widehat{\mathfrak P}_{\mathrm{bal}}(Y)
=
\argmin_{\mathcal R\in\Pi_{n,2}^{\mathrm{bal}}}
F_Y(\mathcal R).
\]
This set is nonempty because $\Pi_{n,2}^{\mathrm{bal}}$ is finite.
When the balanced optimum is unique, denote its sole element by
$\widehat{\mathcal P}_{\mathrm{bal}}(Y)$.

For two $n\times p$ matrices $Y$ and $\widetilde Y$ with rows
$\mathbf y_i^\top$ and $\widetilde{\mathbf y}_i^\top$, respectively, let
\begin{align}\label{def:N.num.dif.rows}
N_R(Y,\widetilde Y)
=
\#\{i\in[n]:\mathbf y_i\ne\widetilde{\mathbf y}_i\},
\end{align}
the number of row indices on which they differ. Define the balanced
rowwise half-Rand corruption count by
\begin{equation}
B_{R,\mathrm{bal}}^{(1/2)}(X)
=
\min_{\widetilde X}
\left\{
N_R(X,\widetilde X):
\begin{array}{l}
\widehat{\mathfrak P}_{\mathrm{bal}}(\widetilde X)=\{\mathcal Q\}
\text{ for some }\mathcal Q\in\Pi_{n,2}^{\mathrm{bal}},\\[1mm]
d_{\mathrm R}(\mathcal P,\mathcal Q)\ge\frac12
\end{array}
\right\}.
\label{eq:balanced-half-rand-count}
\end{equation}

\begin{proposition}[Balanced rowwise half-Rand robustness]
\label{prop:balanced-row-half-rand}
Let
\begin{equation}
t_n^{\mathrm{bal}}
=
\min\left\{
s\in2\mathbb Z:
0\le s\le n/2,\quad
\frac{s(n-s)}{\binom n2}\ge\frac12
\right\}.
\label{eq:balanced-half-rand-threshold}
\end{equation}
For the benchmark matrix \eqref{eq:balanced-half-rand-benchmark},
\begin{equation}
B_{R,\mathrm{bal}}^{(1/2)}(X)
=
t_n^{\mathrm{bal}},
\qquad
\frac{B_{R,\mathrm{bal}}^{(1/2)}(X)}{n}
\longrightarrow
\frac12
\qquad\text{as }n\to\infty.
\label{eq:balanced-half-rand-exact}
\end{equation}
\end{proposition}

Thus, in this balanced benchmark, nearly half of the observational
units must be altered before the unique optimal balanced partition can
disagree with the original partition on at least half of all pairwise
co-membership decisions.

More generally, if $B_{R,\mathrm{bal}}^{c}$ denotes the analogous
minimum number of altered rows required so that the unique optimal
balanced partition $\mathcal Q$ satisfies
$d_{\mathrm R}(\mathcal P,\mathcal Q)\ge c$, then
\[
\frac{B_{R,\mathrm{bal}}^{c}}{n}
\longrightarrow
\beta(c),
\qquad
\beta(c)=\frac{1-\sqrt{1-2c}}{2},
\qquad 0\le c\le\frac12, 
\]
where the function $\beta(c)$ is plotted in Figure \ref{fig:beta-half-interval}. The proof, including this extension, is given in the Supplementary Material.

\begin{figure}[htbp]
\centering
\includegraphics[width=0.72\linewidth]{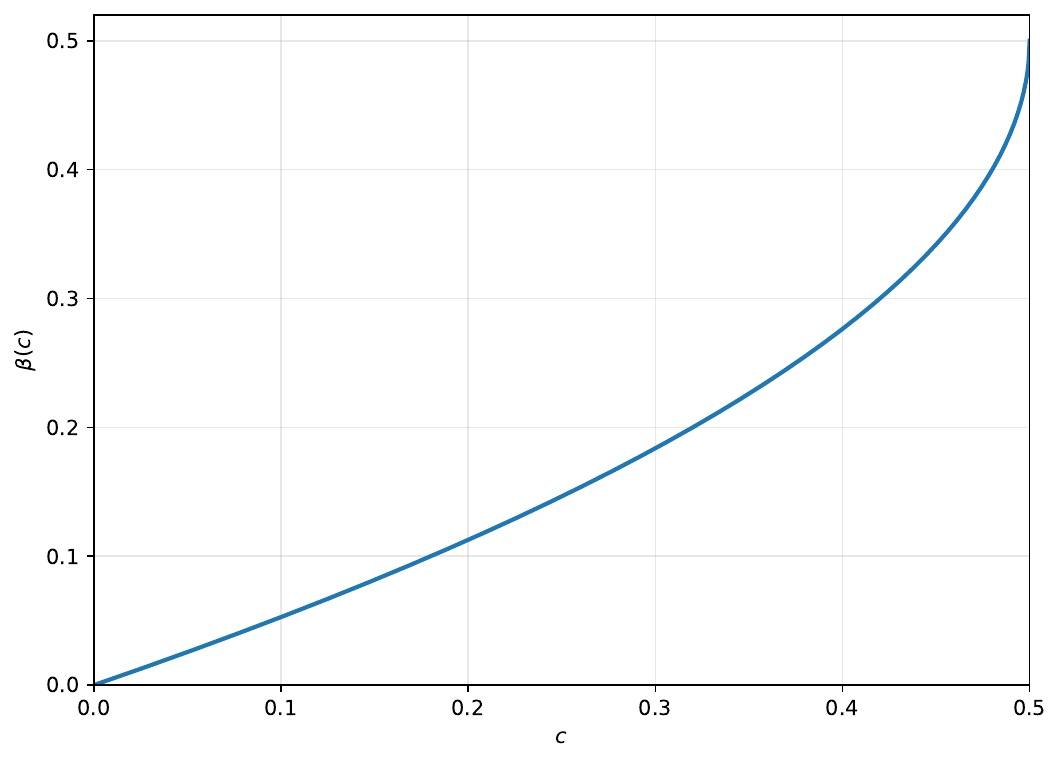}
\caption{The function
$\beta(c)=\frac{1-\sqrt{1-2c}}{2}$ on the admissible range
$0\le c\le 1/2$.}
\label{fig:beta-half-interval}
\end{figure}

For the following complexity bound we assume $O(n \log n)$ and $O(n)$ order for sorting and extracting a median from a list of size $n$, respectively.

\begin{proposition}[Termination and computational cost]
\label{prop:kmr-current-algorithm}
Algorithm~\ref{alg:kmid}
terminates after finitely many steps to a partition which has exactly $K$ non-empty clusters, is a fixed point, and which need not be globally optimal. If $T$ is the number of iterations performed before the fixed point is
reached then the total running time, using the squared Euclidean or $L^1$ loss is 
$O\!\left(pn\log n+TnpK\right)$.
\end{proposition}

\section{Simulation and validation}\label{sec:exp}

\subsection{Comparison with existing procedures}
In this section, we assess the quality of  the clustering produced by $K$-midranks on synthetic data by withholding the known labels using two metrics, Accuracy and the Adjusted Rand Index. As these two measures emphasize somewhat different aspects of agreement,
both measures are reported in the tables and displayed graphically below.

Let $\mathbf{y}=(y_1,\ldots,y_n)$ and $\mathbf{z}=(z_1,\ldots,z_n)$ denote the true labels and the estimated cluster labels, respectively. Since cluster labels are arbitrary up to a permutation, the clustering accuracy is defined as
\bea \label{def:Acc}
Acc=\frac{1}{n}\max_{\pi\in\Pi_K}\sum_{i=1}^n \mathbf{1}\bigl(y_i=\pi(z_i)\bigr),
\ena
where $\Pi_K$ denotes the set of all permutations of $\{1,\ldots,K\}$.

The Adjusted Rand Index, or ARI \citep{hubert1985comparing} evaluates the agreement between the predicted clustering and ground truth labels while adjusting for chance grouping. The ARI is defined as:
\bea
\label{def:ARI}
ARI({\bm A},{\bm C}) = \frac{
    \sum_{ij} \binom{|A_i\cap C_j|}{2} - \left[ \sum_i \binom{|A_i|}{2} \sum_j \binom{|C_j|}{2} \middle/ \binom{n}{2} \right]
    }{
\frac{1}{2} \left[ \sum_i \binom{|A_i|}{2} + \sum_j \binom{|C_j|}{2} \right] - \left[ \sum_i \binom{|A_i|}{2} \sum_j \binom{|C_j|}{2} \middle/ \binom{n}{2} \right]
}
\ena
where ${\bm A} = \{A_1, A_2, \dots, A_k\}$ is the partition into $k$ clusters obtained from the algorithm, ${\bm C} = \{C_1, C_2, \dots, C_K\}$ is the true partition, and $n$ is the total number of data points. The ARI ranges from $-1/2$ to $1$, with $1$ indicating perfect agreement.

The ARI has a useful pairwise interpretation.  For each pair of
observations, two partitions agree if they either place both observations in
the same cluster or place them in different clusters.  The ordinary Rand
index \citep{Rand1971} measures this pairwise agreement, while the adjusted Rand index removes
the amount of agreement that would be expected by chance given the cluster
sizes.  Thus an ARI of $0$ means agreement at the chance level, rather than
that the two partitions agree on no pairs; a negative ARI indicates agreement
below the chance level.

We compare the experimental accuracy {\rm Acc} \eqref{def:Acc} and the
Adjusted Rand Index, {\rm ARI} \eqref{def:ARI}, of our $K$-midranks
procedure using the $L^1$ distance (denoted KMR.L1 in the tables and
figures below) and the squared-Euclidean loss (KMR.L2) with those of $K$-means
(Kmeans), $K$-medoids (Kmed), Gaussian mixture models (Gauss),
Student's $t$ mixture models (t), Cauchy mixture models (Cauchy), and
generalized hyperbolic mixture models (G-Hpbl) across datasets from
the four classes mentioned earlier. $K$-medians is not included in our
experiment since it is closely related to the $L^1$ $K$-medoids.

We conduct experiments on four classes of synthetic datasets consisting of $K$ clusters whose labels are withheld. Data is initially generated independently via the standard normal distribution. The datasets are categorized into four classes according to the type of contamination introduced.

In more detail, in the first three classes we create $K \in \{2,\ldots,9\}$ clusters, each containing 1000 individuals, by initially generating feature vectors of length $2K$ for each individual having independent standard normal entries.

\begin{list}{}{} \label{datagroup}
\item {\bf Class 1:} Clusters $i = 1,2,...,K$ are created from the initial configuration by adding independent normal random variables in coordinates $2i-1$ and $2i$ with mean $sepp \in \{1.5,2\}$ and standard deviation $1$.  

\item {\bf Class 2:} We proceed as for Class 1, but using the Student's $t$ distribution with location parameter $sepp \in \{1.5,2\}$ and $3$ degrees of freedom.

\item {\bf Class 3:} As for Class 1, but the shifts follow a Cauchy distribution with location parameter $sepp \in \{1.5,2\}$ and scale parameter $0.3$.

\item {\bf Class 4:}
We generate data as for Class 3, but for clusters of different sizes, with $K_1 \in \{2,3,4\}$ clusters that contain 1000 individuals, and $K_2 \in \{2,3,4\}$ clusters that contain 2000 individuals, for a total of $K=K_1+K_2$ clusters. In addition, we perform the Cauchy shifts as for Class 3, with the number of shifted coordinates independently taking values in $\{2,3\}$ for each group of clusters.

\end{list}

All tables and figures presenting the results are collected in Appendix~\ref{appexdix:sim}. The results are discussed separately for each data class in the subsections below.

\subsubsection{Class 1 result discussion}

Class 1 examines the performance cost of replacing numerical magnitudes
by ranks when the data are Gaussian. As shown in
Table~\ref{tab:class1} and Figures~\ref{fig:gau} and~\ref{fig:gau2},
the Gaussian mixture model performs best according to both accuracy
and ARI in 13 of the 16 settings. Other mixture models perform best
in the remaining three settings, and $K$-means generally performs
nearly as well as the best-performing method.

The performance disadvantage of $K$-midranks varies considerably across
settings. At $K=2$ and separation $2$, the accuracies of its $L^1$ and
squared-Euclidean versions are $0.945$ and $0.941$, compared with
$0.956$ for the Gaussian mixture model. At $K=9$ and separation $1.5$,
however, the corresponding accuracies are $0.363$ and $0.529$, compared
with $0.687$ for the Gaussian mixture model; the ARIs are $0.099$ and
$0.223$, compared with $0.417$.

Thus, on these Gaussian data, $K$-midranks performs worse than Gaussian
mixture clustering and $K$-means. The difference is small in some
settings with few clusters and larger separation, but can be
substantial in the larger, lower-separation configurations.

\subsubsection{Class 2 result discussion}

Class 2 uses Student's $t$ shifts with three degrees of freedom.
Table~\ref{tab:simpt} and Figures~\ref{fig:t} and~\ref{fig:t2} show
that the best-performing mixture models outperform both $K$-midranks
versions in every displayed setting. $K$-means also usually outperforms
both versions. The Cauchy mixture most often performs best at
separation $1.5$, while Gaussian and Student's $t$ mixtures frequently
perform best at separation $2$.

Not every mixture model outperforms every representative-based method.
For example, at $K=3$ and separation $2$, the Gaussian mixture performs
worse than $K$-means on both measures: its accuracy and ARI are $0.623$
and $0.433$, compared with $0.882$ and $0.677$ for $K$-means. The
Student's $t$ mixture performs better than both, with accuracy $0.911$
and ARI $0.751$. The performance disadvantage of $K$-midranks can also
be substantial: at $K=9$ and separation $1.5$, its $L^1$ and
squared-Euclidean ARIs are $0.058$ and $0.190$, compared with $0.374$
for $K$-means and $0.397$ for the Cauchy mixture.

These results show that the presence of heavy tails alone does not
ensure an advantage for rank-based clustering. In the Student's
$t$-shifted settings considered here, retaining the numerical
magnitudes gives better results for the leading mixture models and,
in most settings, for $K$-means.

\subsubsection{Class 3 result discussion}

Replacing the Student's $t$ shifts by Cauchy shifts changes the
comparison markedly. Table~\ref{tab:simpc} and
Figures~\ref{fig:cau} and~\ref{fig:cau2} show that, among methods
with reported results, one of the two $K$-midranks versions performs
best according to both accuracy and ARI in each of the 14 settings
with $K\ge3$. At $K=2$, the generalized hyperbolic mixture performs
best at separation $1.5$, whereas the Cauchy mixture performs best
at separation $2$.

The choice of loss matters. At separation $1.5$, the $L^1$ version
performs better than the squared-Euclidean version according to ARI
for $K=2,\ldots,5$, while the squared-Euclidean version performs
better according to both accuracy and ARI for $K=6,\ldots,9$.
At $K=9$, for example, the accuracies of the $L^1$ and
squared-Euclidean versions are $0.294$ and $0.566$, and their ARIs
are $0.081$ and $0.263$, respectively. At separation $2$, the $L^1$
version performs better on both measures through $K=8$, with the
squared-Euclidean version performing better at $K=9$. Thus a
performance advantage for one $K$-midranks version does not imply
that both versions outperform all competitors.

The effect of separation on the performance advantage over competitors
depends on the setting and the performance measure. At $K=9$ and
separation $2$, squared-Euclidean $K$-midranks performs only slightly
better than the Gaussian mixture: their accuracies are $0.730$ and
$0.719$, and their ARIs are $0.484$ and $0.482$. The generalized
hyperbolic mixture is reported as not converging in seven settings,
so comparisons in those settings are limited to methods with reported
results.

In these experiments, Cauchy shifts provide a clear setting in which
rank transformation improves clustering performance. Neither loss is
uniformly preferable, and the squared-Euclidean version is particularly
advantageous in the larger-$K$, lower-separation configurations.

\subsubsection{Class 4 result discussion}

Class 4 combines Cauchy shifts with unequal cluster sizes and
heterogeneous shift configurations. Table~\ref{tab:2diff1} reports
18 settings, and Figure~\ref{fig:class4} displays their results
ordered by decreasing ARI for $L^1$ $K$-midranks. The
squared-Euclidean version of $K$-midranks outperforms every
non-$K$-midranks competitor with reported results on both accuracy
and ARI in all 18 settings; the $L^1$ version does so in 17.
In the direct comparison between the two rank-based versions,
$L^1$ performs better on both measures in ten settings, while
squared Euclidean performs better in seven. They tie in one setting
at the reported precision.

The size of the performance advantage varies. For example, with
$K_1=3$, $K_2=4$, and separation parameters $1.5$ and $2$,
squared-Euclidean $K$-midranks performs better than the Gaussian
mixture on both measures: its accuracy and ARI are $0.717$ and
$0.487$, compared with $0.642$ and $0.454$ for the Gaussian
mixture. In this setting, however, the $L^1$ version performs worse
than the Gaussian mixture, with accuracy $0.626$ and ARI $0.406$.
The generalized hyperbolic mixture is reported as not converging
in eleven settings. Our comparisons concern the displayed settings
and the results available for them.

Thus both versions of $K$-midranks generally outperform the other
clustering methods in these experiments with unequal cluster sizes
and heterogeneous Cauchy shifts. Although the $L^1$ version more
often performs better in the direct comparison between the two
rank-based versions, the squared-Euclidean version outperforms all
non-$K$-midranks competitors with reported results in every displayed
setting.

\subsection{Sensitivity analysis}\label{subsec:sensitivity}

We assess the sensitivity of the $K$-midranks algorithm to initialization using heavy-tailed datasets from Class 3, the setting for which the algorithm was designed. For each dataset ($K=2, 5,$ or $9$), we run the algorithm 30 times with different random initializations. Table \ref{tab:sens} reports the mean, standard deviation (SD), minimum, and maximum of both clustering accuracy and the adjusted Rand index (ARI) across the 30 runs. For $L^1$ $K$-midranks, the results are relatively stable across initializations for $K=2$ and $K=5$ at both separations. However, greater variability is observed for $K=9$, particularly at a separation of $1.5$, where the mean accuracy is $0.322$ (SD $0.024$) and the mean ARI is $0.092$ (SD $0.010$). At a separation of $2$, the mean accuracy increases to $0.723$ (SD $0.039$), while the mean ARI increases to $0.480$ (SD $0.040$). In comparison, squared-Euclidean $K$-midranks exhibits smaller variability at a separation of $2$, with SDs of $0.001$ for both accuracy and ARI when $K=9$.

\subsection{Detecting the correct number of clusters}
In practice, the true number of clusters will be unknown to the user.
We investigate its selection using the classical elbow principle
\cite{Thorndike1953}.

For a fitted partition into $k$ clusters,
${\bm C}^{(k)}=\{C_1^{(k)},\ldots,C_k^{(k)}\}$, let
\[
D_q(k)
=
\sum_{j=1}^k
\sum_{i\in C_j^{(k)}}
d_q^q\!\left(\mathbf R_i,\mathbf v_j^{(k)}\right),
\]
where $\mathbf v_j^{(k)}$ is the representative of cluster
$C_j^{(k)}$.  Thus $D_q(k)$ is the total within-cluster dispersion
for the $k$-cluster solution, with smaller values corresponding to
tighter clusters.

Following the one-step elbow index of \citet{RykovEtAl2024}, define,
for $k=2,3,\ldots$,
\bea \label{ebformula}
B_q(k)
=
\frac{D_q(k-1)-D_q(k)}
     {D_q(k)-D_q(k+1)}.
\ena
The numerator measures the reduction in within-cluster dispersion
obtained by increasing the number of clusters from $k-1$ to $k$,
whereas the denominator measures the additional reduction obtained
by increasing it from $k$ to $k+1$. 
Hence a value of $B_q(k)$ substantially larger than $1$ indicates that
the reduction in within-cluster dispersion obtained by increasing the
number of clusters from $k-1$ to $k$ is appreciably greater than the
additional reduction obtained by increasing the number from $k$ to
$k+1$. 
This is precisely the
flattening of the dispersion curve that the elbow method seeks to
detect.  We therefore select the value of $k$ for which $B_q(k)$ is
largest.

We report results for the datasets from Class 3 in Table \ref{tab:elb1}, which reports the values of $k$ corresponding to the two largest values of $B_q(k)$. At a separation of $2$, the criterion in \eqref{ebformula} correctly identifies the true number of clusters $K$ in all settings using the $L^1$ distance. For the squared-Euclidean distance, the criterion correctly identifies $K$ for $K \leq 8$, but selects $2$ when $K=9$, with the true value $9$ ranked second. For datasets with a separation of $1.5$, however, it fails to identify the correct number of clusters when $K \ge 7$.

\subsection{Run time}

Table~\ref{tab:computational-results} gives runtimes for the Class~3
datasets, which we see are consistent with the theoretical computational
complexity bound established in Proposition~\ref{prop:kmr-current-algorithm}. The (scaled up by $10^9$) ratio of the runtimes to the complexity expression $pn\log n+TnpK$, from Proposition~\ref{prop:kmr-current-algorithm}, appears to stabilize for the $L^1$ case around $13$ for $n\ge 8000$, and to 
$10$ for $n\ge 5000$, for squared Euclidean loss.

Figure~\ref{fig:runtime-complexity} further illustrates this relationship. For both the $L^1$ and $L^2$ cases, we see that the observed runtimes exhibit an approximately linear relationship with the complexity expression given.
Although deviations from the fitted lines are observed, particularly for the $L^1$ algorithm, the overall trends are consistent with the theoretical complexity bound over the ranges considered.

All experiments were conducted on a laptop equipped with an Intel Core Ultra 7 165H (1.40 GHz), 64 GB RAM, running Windows 11. The algorithms were implemented in R 4.4.1  using R packages `clue', `cluster', `factoextra', `mclust', `teigen', and `MixGHD' through RStudio.  Our R package for $K$-midranks, called `kmidrank' is currently available at \url{https://github.com/nwiroonsri/kmidranks}.

\section{Real-data analyses and substantive findings}\label{sec:app}

In this section, we apply $K$-midranks to three real-world datasets including Raman and LIBS spectroscopy, and S\&P500 returns.

Before interpreting the clustering results, we examine whether the
real datasets display distributional behavior for which a rank-based
procedure may be advantageous. Figure~\ref{fig:qq1} provides benchmarks
from the simulated datasets, ranging from the approximately Gaussian
behavior of Class~1 through the heavier-tailed Student's $t$ setting of
Class~2 to the Cauchy-based Classes~3 and~4. The corresponding Q--Q plots
for the real datasets are shown in Figure~\ref{fig:qq2}. Several dimensions
of the LIBS and S\&P 500 data show substantial outward departures in both
tails from the normal reference line, indicating the presence of observations
far more extreme than would be expected under a Gaussian model. The Raman
data also depart markedly from normality, although their Q--Q plots have a
different shape, with pronounced changes in slope rather than the pattern observed in the Cauchy simulations.  These diagnostics provide
empirical motivation for applying a procedure which is robust to heavy tailed data.

\subsection{Raman spectroscopy}\label{subsec:raman}

Raman spectroscopy is a form of spectroscopy distinguished by its use of
light scattered from a sample rather than light absorbed or emitted by it.
A monochromatic beam, usually produced by a laser, is directed at the
sample.  Almost all of the scattered light has the same frequency as the
incident beam, but a very small fraction emerges with a slightly different
frequency because energy has been exchanged between the light and molecular
vibrations.  The resulting frequency changes, called Raman shifts, depend on
the bonds and molecular structure of the substance and therefore produce a
characteristic fingerprint.  In this respect Raman spectroscopy is
closely related to infrared spectroscopy, which also probes molecular
vibrations, but infrared spectroscopy measures the absorption of radiation,
whereas Raman spectroscopy measures these frequency shifts in scattered
light.  The two techniques consequently provide complementary information
about molecular structure.

Because Raman measurements can be rapid and
nondestructive, they are used in areas such as mineral identification,
pharmaceutical analysis, and the monitoring of chemical and crystallization
processes.  For example, Raman spectra can distinguish different crystalline
forms, or polymorphs, of the same pharmaceutical compound, which is important
because different polymorphs of an active pharmaceutical ingredient can have
different physical and therapeutic properties.

The dataset of \citet{flanagan2025open} was assembled to address a
practical limitation in pharmaceutical research: despite the extensive use
of Raman spectroscopy in drug discovery and active
pharmaceutical ingredient (API) development, relatively few large,
high-quality Raman spectral libraries are openly available.  The authors
therefore collected spectra from 32 commercial chemical substances,
primarily solvents and reagents used in API development.  These substances
were not chosen arbitrarily: only products that had been tested and
successfully used in API-development work were included, where individual
chemicals may serve, for instance, as solvents or
reactants.  The resulting dataset contains 3,510 spectra and was intended
to provide reference data for spectral analysis.

Although the dataset was not created specifically for clustering, it
provides a natural setting in which to study whether an unsupervised clustering method
can recover chemically meaningful structure from Raman measurements alone.
Spectra from the same substance should exhibit a common molecular signature,
while spectra from different substances should in principle be
distinguishable through differences in their Raman features. In our study the known identities of the 32
substances are withheld during clustering and used only afterward as
external labels for evaluating how well the recovered groups correspond to
the underlying chemical classes.

To reduce the original dimension from 3,276, we apply PCA and use the first 20 components, which explain about 0.999999 of the variance. 
Table~\ref{tab:rw} shows that $K$-midranks with $L^1$ and $L^2$, Gaussian and Cauchy mixture models are among the best performers, with accuracies of 0.829, 0.759, 0.789, and 0.681, and ARI values of 0.733, 0.698, 0.656, and 0.664, respectively. In comparison, $K$-means achieves an accuracy of 0.499 and an ARI of 0.354, while $K$-medoids achieves an accuracy of 0.569 and an ARI of 0.407. Student's t and generalized hyperbolic mixture models did not converge. Figure \ref{fig:raman} shows all 32 groups on the first two principal components. We note the strong similarity between the true and clustering labels, reflecting the accuracy of 0.829 and the ARI of 0.733.

Figure~\ref{fig:raman2} shows the frequencies of true and false detections for each of the 32 materials using $K$-midranks with $L^1$ distance. These results reveal substantial variation in the ability of $K$-midranks to distinguish different Raman spectra. Isobutylamine, Diethylamine, and Formic acid are perfectly detected, with all their observations assigned to their corresponding matched clusters. In contrast, 2,2-Dimethoxypropane, Dimethyl sulfoxide, and Dichloromethane have detection rates of only 0\%, 21\%, and 46.8\%, respectively, thus appearing to be considerably more difficult for $K$-midranks to distinguish. Interestingly, the poor detection of 2,2-Dimethoxypropane is highly systematic rather than being caused by assignments across several different materials. Among its 107 observations, 100 are grouped with tert-butyl methyl ether. This suggests that the two materials are difficult to distinguish using $K$-midranks with $L^1$ distance, possibly because their Raman spectra exhibit similar rank patterns across the retained principal component scores.

\subsection{Laser-induced breakdown spectroscopy}\label{subsec:libs}


Laser-induced breakdown spectroscopy (LIBS) is an emission spectroscopy
technique used primarily to determine the elemental composition of a
material.  A short, high-powered laser pulse is focused onto a very small
region of the sample, where the deposited energy vaporizes and ionizes a
tiny amount of material and creates a short-lived plasma.  The word
``breakdown'' in the name refers to this laser-induced transition to an
ionized plasma.  As the
plasma cools, its atoms and ions emit light at characteristic wavelengths.
Recording the intensities of these emission lines therefore produces a
spectral signature from which the elements present in the material can be
inferred.

LIBS is particularly useful when rapid measurements must be made directly
on a material with little sample preparation, and it has consequently been
used in applications including geological analysis and industrial materials
identification \citep{kepevs2020benchmark}.  These
applications naturally give rise to clustering and classification problems:
samples having similar elemental compositions should tend to produce
similar emission spectra, while distinct material classes should exhibit
different spectral patterns.

The dataset of \citet{kepevs2020benchmark} was created specifically as a
benchmark for developing and evaluating LIBS classification and clustering methods. It contains spectra from 138 prepared geological samples belonging
to 12 ore and geological material classes, numbered 1--12 in
Table~\ref{tab:libs-classes}. The dataset is divided into training and test subsets. For our illustration we use only the training subset, which contains 100 specimens. We select the first 100 spectra from each specimen, resulting in 10,000 spectral observations. Each recorded spectrum consists of emission intensities measured over a large number of wavelengths and is therefore represented as a high-dimensional feature vector. The benchmark dataset was designed to provide a more realistic classification and clustering problem than simply distinguishing repeated measurements of the same specimen. Specifically, different physical samples belong to the same geological class, and additional mixing was used to introduce controlled within-class variation.  Thus a successful method must identify the common
spectral structure of a geological class rather than merely recognize a
particular sample.  In our unsupervised setting, the class labels are withheld
from the clustering procedure and used afterward to assess whether the groups
recovered from the LIBS spectra correspond to the known geological classes.
There are 12 such classes, and we use $K=12$ in this analysis.

Since the data is high-dimensional, we again apply PCA and use the first 50 principal components as inputs for clustering. These components explain 0.999999 of the total variance. None of the approaches performs particularly well on this dataset. Nevertheless, $L^1$ $K$-midranks achieves the highest accuracy ($Acc = 0.536$, $ARI = 0.303$), followed closely by $L^2$ $K$-midranks ($Acc = 0.526$, $ARI = 0.306$). The generalized hyperbolic mixture model achieves the highest ARI ($Acc = 0.504$, $ARI = 0.343$). The other mixture models yield slightly lower accuracy and ARI, while $K$-means and $K$-medoids perform particularly poorly on the LIBS dataset. Figure~\ref{fig:LIBs} displays all 12 groups projected onto the first two principal components using true labels and $K$-midranks labels. We note that the true and the clustering labels are not closely aligned, reflecting the accuracy of 0.536 and the ARI of 0.303.

The heatmap in Figure~\ref{fig:LIBs2} provides a more detailed view of the class-specific clustering performance of $K$-midranks. Classes 10 and 12 are the most successfully detected, with 95.2\% and 87.0\% of their observations assigned to their corresponding matched clusters, respectively. In contrast, Classes 3, 4, and 8 are among the most difficult to recover. The misclassifications of these classes exhibit relatively structured patterns. For Class 3, 36.0\% of observations are assigned to Cluster 3, while substantial proportions are assigned to Clusters 4 (20.4\%), 7 (18.8\%), and 1 (14.4\%). For Class 4, only 33.8\% of observations are assigned to Cluster 4, with substantial proportions assigned to Clusters 10 (30.0\%), 5 (14.2\%), and 1 (11.7\%). Class 8 has 50.0\% of observations assigned to Cluster 4 and 32.2\% to Cluster 8, indicating considerable misclassification. These results show considerable variation in the ability of $K$-midranks to recover the predefined LIBS classes. Given the overall clustering accuracy of approximately 0.50, the observed misclassification patterns should be interpreted cautiously and do not by themselves establish similarities in the underlying chemical compositions of these classes.

\subsection{S\&P 500 returns}

The S\&P 500 dataset from Kaggle \citep{larxel2021sp500} contains stock information for 751 trading days from 2023 to 2025 for companies included in the S\&P 500 index, one of the most widely used benchmarks for the U.S. equity market. For clustering, each company is represented by its daily log returns over the entire observation period, and the clustering algorithms are applied to the complete dataset to obtain a single partition of the companies.

The primary objective is to perform unsupervised clustering of stocks based solely on their log return patterns, without imposing predefined sector classifications. However, since companies within the same sector often share common economic drivers and exhibit similar market behaviors, sector labels can serve as useful external benchmarks for evaluating the quality and interpretability of the clusters obtained. We conduct experiments using three, five, and eight sectors to examine whether the clustering methods can identify groups of stocks that correspond to known sector classifications or reveal similarities across different sectors. Since one objective is to use business sectors as benchmark labels, we deliberately select sectors with distinct business activities, as listed below.

\begin{itemize}
    \item Three sectors: Communication Services, Utilities, and Real Estate
    \item Five sectors: Information Technology,  Health Care, Energy, Consumer Staples, and Real Estate
    \item Eight sectors: Consumer Staples, Financials, Industrials, Energy, Health Care, Materials, Real Estate, and Utilities
\end{itemize}

The eight sector selection is the same as in \cite{javaheri2023graph}. However, these authors evaluated their results only on a selected group of 100 out of 352 companies from the eight sectors, and they used 1000 days of daily returns. Moreover, their proposed method was designed for time-varying data while $K$-midranks is for general clustering.

Table \ref{tab:rw} shows that $K$-midranks achieves the best clustering performance on the S\&P500 data among the methods considered. For the three- and five-sector cases, $L^2$ $K$-midranks achieves the highest accuracy (0.929 and 0.818) and ARI (0.802 and 0.633), respectively. For the eight-sector case, $L^1$ $K$-midranks achieves the highest accuracy (0.691) and ARI (0.507). We note that replacing log-returns with simple returns yields identical results for $K$-midranks, since the strictly increasing transformation $\log(1+r)$ preserves ranks and hence leaves the rank-space representation unchanged as stated in Proposition \ref{prop:kmr-current-invariance}.


As mentioned earlier, our goal is to identify groups of companies with similar patterns and behaviors, regardless of their business sectors. To illustrate how the clusters identified by $K$-midranks relate to the business-sector classification, we examine Cluster 7 from the eight-sector case, as shown in Figure~\ref{fig:sp}.  Figure~\ref{fig:sphm} presents a heatmap showing the correspondence between the clustering labels and business sectors. Similar analyses can be conducted for the remaining clusters and for the three and five sector cases, but we restrict attention to this example for brevity.

The left panel of Figure~\ref{fig:sp} shows the mean daily closing-price trajectories of the eight clusters obtained using $L^1$ $K$-midranks, with Cluster 7 highlighted in red. This cluster exhibits a comparatively flat mean trajectory over the observation period, whereas several other clusters display more pronounced upward trends. The middle panel displays individual trajectories within Cluster 7, colored by sector, together with the cluster mean in red. For visual clarity, companies with particularly high closing prices are omitted from this panel to make the remaining trajectories easier to distinguish; however, the cluster mean is calculated using all companies in the cluster. Cluster 7 consists primarily of Real Estate companies but also includes companies from Consumer Staples, Financials, Health Care, Industrials, and Materials. Although the displayed trajectories vary in price level and short-term fluctuations, many exhibit relatively limited overall growth.

The right panel of Figure~\ref{fig:sp} shows the sector-wide mean trajectories of the six sectors represented in Cluster 1. In particular, the Financials and Industrials sector means show clearer upward trends than the mean of Cluster 7. These comparisons suggest that $K$-midranks identifies a grouping across sector boundaries, bringing together companies whose price trajectories can differ from the aggregate behavior of their respective sectors.

This example illustrates the potential of $K$-midranks to identify groupings of companies across sector boundaries based on their stock behavior and hence to potentially provide additional insights beyond the existing sector classification.

\section{Discussion and limitations}\label{sec:conclusion}

\subsection{When to Use $K$-midranks}

Our experiments suggest that $K$-midranks is particularly effective for data with very heavy tails, as illustrated by the Cauchy-based simulations, where the underlying distributions do not have finite means. In contrast, $K$-means and model-based methods perform better in the Gaussian and Student's $t$ settings considered here, including the heavy-tailed $t$ distribution with three degrees of freedom. Thus, heavy tails alone do not imply that $K$-midranks is to be preferred. However, we note that on real data, $K$-midranks performs slightly better than the model-based methods on the spectroscopy datasets and achieves a more substantial advantage on the S\&P500 data. These findings suggest that $K$-midranks is a useful candidate when extreme observations may adversely affect methods that rely on the magnitudes of the original measurements.

To investigate whether heavy-tailed distributions may help explain the experimental results across the simulated and real-world datasets, we constructed normal Q--Q plots for individual variables, each standardized by its sample mean and standard deviation. Each plot uses observations from a single variable pooled across all clusters.

For the simulated datasets, we select one representative case from each class for illustration. Specifically, we consider $K=9$ and $\mathrm{sepp}=1.5$ for Classes 1--3, and $K_1=K_2=4$ and $\mathrm{sepp}_1=\mathrm{sepp}_2=1.5$ for Class 4. Figure~\ref{fig:qq1} presents the normal Q--Q plots for all variables in each selected dataset. The Gaussian-based data in Class 1 follow the normal reference line closely, whereas the $t$-based data in Class 2 exhibit pronounced departures in both tails. The Cauchy-based data in Classes 3 and 4 display much flatter central portions and substantially more extreme tail departures than those observed in Class 2. These differences illustrate the distinct tail behaviors among the four simulation classes.

For the real-world datasets, we examined the 20 and 50 retained principal components of the Raman and LIBS data, respectively, and 15 randomly selected trading dates from the S\&P 500 dataset, each representing the cross-sectional distribution of returns across companies. As shown in Figure~\ref{fig:qq2}, LIBS and S\&P500 exhibit pronounced departures from the normal reference line in both tails across several variables, with the lower tails falling below and the upper tails rising above the reference line. These patterns suggest heavy-tailed behavior and are similar to those observed in the $t$-based simulations of Class 2, although the magnitude of the tail departures varies across variables. In contrast, the Raman data exhibit less consistent evidence of heavy-tailed behavior, with pronounced changes in slope and flattening over portions of the Q--Q curves rather than the more systematic tail departures observed in LIBS and S\&P500. We note that these observations provide only exploratory evidence of heavy-tailed behavior.

\subsection{Conclusion and future work}
In this work, we introduce a new clustering method, called $K$-midranks, which replaces raw data with coordinatewise ranks for each variable and then handles them similarly to the classic $K$-means. This transformation is intended to reduce the influence of outliers and heavy tails. 

$K$-midranks is evaluated and compared with existing clustering algorithms, including $K$-means, $K$-medoids, and model-based clustering based on Gaussian, Student's t, Cauchy, and generalized hyperbolic distributions, using simulated datasets. It is then applied to several real-world applications, including Raman and LIBS spectroscopy and S\&P 500 stock returns. Its performance on these real-world datasets is evaluated using the available external labels. In the spectroscopy applications, $L^1$ $K$-midranks achieves the highest clustering accuracy, although model-based methods remain competitive in terms of ARI. $K$-midranks also achieves the best performance on S\&P 500 returns, which are well known to exhibit heavy-tailed behavior.

Despite these promising results, several limitations of the proposed method should be acknowledged. $K$-midranks underperforms model-based clustering on Student's t data, which also have heavy tails but with a finite mean. In practice, it would be hard to know when a heavy tailed distribution lacks a first moment. Compared to $K$-means, $K$-midranks has the additional computational burden of ordering the observations and replacing them with ranks.

Our R package for $K$-midranks, currently available at \url{https://github.com/nwiroonsri/kmidranks}, makes this method practically accessible. Future work includes exploring more situations in which $K$-midranks is suitable, further investigating cluster validation techniques for $K$-midranks, and applying $K$-midranks to a broader range of real-world problems.

\begin{acks}[Acknowledgments]
The second author would like to thank King Mongkut's University of Technology Thonburi for providing a Skill Development Grant to support his research visit to the University of Southern California.
\end{acks}

\begin{funding}
The second author was supported in part by the Skill Development Grant 2026, King Mongkut's University of Technology Thonburi.
\end{funding}

\section*{Statement on AI use}

Generative AI tools were used to assist in drafting and refining portions of the manuscript, improving language and readability, and identifying potential inconsistencies and ambiguities in the presentation. AI was also used to assist in checking mathematical proofs, verifying and debugging computational code, and supporting real-data analyses. The research ideas, proposed methodology, and mathematical results were developed by the authors. All AI-assisted content, mathematical arguments, computational implementations, and numerical results were reviewed and verified by the authors, who take full responsibility for the accuracy and integrity of the manuscript.

\begin{supplement}
Proofs of the mathematical properties stated in the main paper,
together with additional theoretical results appear in the accompanying supplement. 
\end{supplement}

\appendix

\section{Simulation and validation results} \label{appexdix:sim}

This section presents tables and plots that summarize the simulation and validation results. These include comparisons of clustering performance, sensitivity analyses, estimates of the number of clusters, and computational performance (runtime).

\subsection{Clustering performance tables}

The following tables compare the clustering performance of the methods across the simulated data classes and parameter settings.

\begin{table}[H] 
\centering
\footnotesize
\begin{threeparttable}
\caption{Class 1: Gaussian data. Accuracy and ARI for each dataset in this class. The highest Accuracy and ARI values in each row are shown in bold. Here, $K$ is the number of clusters, $n$ is the total sample size, $p$ is the number of variables, and $\mathrm{sepp}$ is the separation parameter. The same notation applies to Classes 2 and 3.}
\label{tab:class1}
\begin{tabular}{cccccccccccc}
\toprule
  &      &    &    & \multicolumn{2}{c}{Kmeans} &
\multicolumn{2}{c}{Kmed} &
\multicolumn{2}{c}{KMR.L1} &
\multicolumn{2}{c}{KMR.L2}      \\
K & n    & p  & sepp & Acc    & ARI   & Acc   & ARI   & Acc    & ARI   & Acc    & ARI    \\
2 & 2000 & 4  & 1.5  & 0.899      & 0.637      & 0.882    & 0.583    & 0.877      & 0.57       & 0.872      & 0.552      \\
3 & 3000 & 6  & 1.5  & 0.819      & 0.53       & 0.792    & 0.472    & 0.788      & 0.465      & 0.776      & 0.44       \\
4 & 4000 & 8  & 1.5  & 0.802      & 0.542      & 0.756    & 0.453    & 0.755      & 0.453      & 0.736      & 0.419      \\
5 & 5000 & 10 & 1.5  & 0.757      & 0.484      & 0.659    & 0.329    & 0.653      & 0.325      & 0.659      & 0.33       \\
6 & 6000 & 12 & 1.5  & 0.734      & 0.463      & 0.582    & 0.255    & 0.622      & 0.299      & 0.626      & 0.304      \\
7 & 7000 & 14 & 1.5  & 0.7        & 0.422      & 0.533    & 0.216    & 0.448      & 0.149      & 0.581      & 0.263      \\
8 & 8000 & 16 & 1.5  & 0.688      & 0.414      & 0.482    & 0.178    & 0.33       & 0.101      & 0.556      & 0.244      \\
9 & 9000 & 18 & 1.5  & 0.672      & 0.397      & 0.349    & 0.106    & 0.363      & 0.099      & 0.529      & 0.223      \\
2 & 2000 & 4  & 2    & 0.95       & 0.808      & 0.947    & 0.799    & 0.945      & 0.792      & 0.941      & 0.778      \\
3 & 3000 & 6  & 2    & 0.913      & 0.756      & 0.9      & 0.722    & 0.896      & 0.711      & 0.885      & 0.686      \\
4 & 4000 & 8  & 2    & 0.89       & 0.729      & 0.874    & 0.692    & 0.867      & 0.677      & 0.855      & 0.649      \\
5 & 5000 & 10 & 2    & 0.877      & 0.717      & 0.835    & 0.628    & 0.832      & 0.623      & 0.815      & 0.591      \\
6 & 6000 & 12 & 2    & 0.856      & 0.684      & 0.784    & 0.546    & 0.794      & 0.567      & 0.778      & 0.537      \\
7 & 7000 & 14 & 2    & 0.846      & 0.673      & 0.753    & 0.497    & 0.772      & 0.539      & 0.758      & 0.515      \\
8 & 8000 & 16 & 2    & 0.83       & 0.649      & 0.714    & 0.449    & 0.738      & 0.491      & 0.734      & 0.485      \\
9 & 9000 & 18 & 2    & 0.814      & 0.625      & 0.66     & 0.372    & 0.664      & 0.392      & 0.707      & 0.45        \\
\end{tabular}
\begin{tabular}{cccccccccccc}
\toprule
  &      &    &      &
\multicolumn{2}{c}{Gauss} &
\multicolumn{2}{c}{t} &
\multicolumn{2}{c}{Cauchy} &
\multicolumn{2}{c}{G-Hpbl}       \\
K & n    & p  & sepp  & Acc    & ARI   & Acc   & ARI   & Acc    & ARI   & Acc    & ARI     \\
2 & 2000 & 4  & 1.5 & 0.903 & 0.651 & 0.902 & 0.648 & 0.902 & 0.646 & \textbf{0.905} & \textbf{0.658} \\
3 & 3000 & 6  & 1.5 & \textbf{0.835} & \textbf{0.567} & 0.829 & 0.553 & 0.828 & 0.551 & 0.831 & 0.558 \\
4 & 4000 & 8  & 1.5 & \textbf{0.811} & \textbf{0.559} & 0.805 & 0.546 & 0.803 & 0.544 & 0.805 & 0.548 \\
5 & 5000 & 10 & 1.5 & \textbf{0.770} & \textbf{0.507} & 0.765 & 0.497 & 0.762 & 0.492 & 0.764 & 0.495 \\
6 & 6000 & 12 & 1.5 & \textbf{0.747} & \textbf{0.485} & 0.734 & 0.462 & 0.731 & 0.458 & 0.731 & 0.457 \\
7 & 7000 & 14 & 1.5 & \textbf{0.715} & \textbf{0.446} & 0.601 & 0.353 & 0.692 & 0.410 & 0.686 & 0.401 \\
8 & 8000 & 16 & 1.5 & \textbf{0.704} & \textbf{0.437} & 0.584 & 0.327 & 0.669 & 0.387 & 0.673 & 0.391 \\
9 & 9000 & 18 & 1.5 & \textbf{0.687} & \textbf{0.417} & 0.630 & 0.339 & 0.639 & 0.352 & 0.635 & 0.347 \\
2 & 2000 & 4  & 2 & \textbf{0.956} & \textbf{0.832} & 0.952 & 0.817 & 0.954 & 0.823 & 0.953 & 0.821 \\
3 & 3000 & 6  & 2 & 0.915 & 0.762 & 0.916 & 0.764 & 0.913 & 0.757 & \textbf{0.917} & \textbf{0.766} \\
4 & 4000 & 8  & 2 & 0.892 & 0.733 & 0.890 & 0.729 & \textbf{0.894} & \textbf{0.738} & 0.891 & 0.731 \\
5 & 5000 & 10 & 2 & \textbf{0.880} & \textbf{0.722} & 0.876 & 0.713 & 0.872 & 0.705 & 0.876 & 0.712 \\
6 & 6000 & 12 & 2 & \textbf{0.862} & \textbf{0.696} & 0.854 & 0.680 & 0.851 & 0.673 & 0.853 & 0.677 \\
7 & 7000 & 14 & 2 & \textbf{0.852} & \textbf{0.685} & 0.844 & 0.669 & 0.841 & 0.663 & 0.843 & 0.667 \\
8 & 8000 & 16 & 2 & \textbf{0.839} & \textbf{0.665} & 0.825 & 0.640 & 0.823 & 0.635 & 0.826 & 0.641 \\
9 & 9000 & 18 & 2 & \textbf{0.826} & \textbf{0.647} & 0.717 & 0.540 & 0.810 & 0.618 & 0.807 & 0.612 \\
\bottomrule
\end{tabular}
\end{threeparttable}
\end{table}

\begin{table}[H] 
\centering
\footnotesize
\caption{Class 2: Gaussian data with simple Student's t shifts. Accuracy and ARI for each dataset in this class. The highest Accuracy and ARI values in each row are shown in bold.}
\label{tab:simpt}
\begin{tabular}{cccccccccccc}
\toprule
  &      &    &    & \multicolumn{2}{c}{Kmeans} &
\multicolumn{2}{c}{Kmed} &
\multicolumn{2}{c}{KMR.L1} &
\multicolumn{2}{c}{KMR.L2}      \\
K & n    & p  & sepp & Acc    & ARI   & Acc   & ARI   & Acc    & ARI   & Acc    & ARI    \\
2 & 2000 & 4  & 1.5  & 0.853      & 0.498      & 0.849    & 0.487    & 0.857      & 0.51       & 0.846      & 0.477      \\
3 & 3000 & 6  & 1.5  & 0.778      & 0.445      & 0.759    & 0.406    & 0.758      & 0.406      & 0.735      & 0.364      \\
4 & 4000 & 8  & 1.5  & 0.75       & 0.444      & 0.726    & 0.4      & 0.702      & 0.363      & 0.684      & 0.335      \\
5 & 5000 & 10 & 1.5  & 0.723      & 0.427      & 0.661    & 0.322    & 0.646      & 0.312      & 0.622      & 0.279      \\
6 & 6000 & 12 & 1.5  & 0.703      & 0.413      & 0.626    & 0.296    & 0.583      & 0.252      & 0.594      & 0.263      \\
7 & 7000 & 14 & 1.5  & 0.681      & 0.393      & 0.535    & 0.212    & 0.457      & 0.148      & 0.557      & 0.234      \\
8 & 8000 & 16 & 1.5  & 0.664      & 0.38       & 0.457    & 0.157    & 0.316      & 0.075      & 0.507      & 0.192      \\
9 & 9000 & 18 & 1.5  & 0.655      & 0.374      & 0.316    & 0.088    & 0.275      & 0.058      & 0.497      & 0.19       \\
2 & 2000 & 4  & 2    & 0.915      & 0.689      & 0.913    & 0.682    & 0.915      & 0.689      & 0.913      & 0.684      \\
3 & 3000 & 6  & 2    & 0.882      & 0.677      & 0.878    & 0.666    & 0.866      & 0.638      & 0.852      & 0.606      \\
4 & 4000 & 8  & 2    & 0.86       & 0.661      & 0.807    & 0.541    & 0.834      & 0.606      & 0.82       & 0.577      \\
5 & 5000 & 10 & 2    & 0.831      & 0.622      & 0.793    & 0.544    & 0.799      & 0.561      & 0.773      & 0.513      \\
6 & 6000 & 12 & 2    & 0.817      & 0.607      & 0.753    & 0.491    & 0.759      & 0.506      & 0.732      & 0.461      \\
7 & 7000 & 14 & 2    & 0.805      & 0.597      & 0.723    & 0.461    & 0.722      & 0.457      & 0.71       & 0.438      \\
8 & 8000 & 16 & 2    & 0.789      & 0.575      & 0.632    & 0.328    & 0.684      & 0.41       & 0.684      & 0.408      \\
9 & 9000 & 18 & 2    & 0.778      & 0.562      & 0.615    & 0.298    & 0.592      & 0.297      & 0.658      & 0.379       \\
\end{tabular}
\begin{tabular}{cccccccccccc}
\toprule
  &      &    &      &
\multicolumn{2}{c}{Gauss} &
\multicolumn{2}{c}{t} &
\multicolumn{2}{c}{Cauchy} &
\multicolumn{2}{c}{G-Hpbl}       \\
K & n    & p  & sepp  & Acc    & ARI   & Acc   & ARI   & Acc    & ARI   & Acc    & ARI     \\
2 & 2000 & 4  & 1.5 & 0.867 & 0.539 & 0.879 & 0.576 & \textbf{0.886} & \textbf{0.597} & 0.874 & 0.559 \\
3 & 3000 & 6  & 1.5 & 0.821 & 0.535 & 0.827 & 0.548 & \textbf{0.831} & \textbf{0.557} & 0.821 & 0.535 \\
4 & 4000 & 8  & 1.5 & \textbf{0.800} & \textbf{0.538} & 0.790 & 0.517 & \textbf{0.800} & 0.537 & 0.787 & 0.511 \\
5 & 5000 & 10 & 1.5 & 0.624 & 0.391 & 0.758 & 0.485 & \textbf{0.765} & \textbf{0.499} & 0.758 & 0.486 \\
6 & 6000 & 12 & 1.5 & \textbf{0.742} & \textbf{0.476} & 0.734 & 0.462 & 0.735 & 0.464 & 0.733 & 0.460 \\
7 & 7000 & 14 & 1.5 & 0.618 & 0.370 & 0.698 & 0.417 & \textbf{0.700} & \textbf{0.422} & 0.693 & 0.410 \\
8 & 8000 & 16 & 1.5 & \textbf{0.698} & \textbf{0.428} & 0.676 & 0.396 & 0.608 & 0.366 & 0.673 & 0.392 \\
9 & 9000 & 18 & 1.5 & 0.637 & 0.320 & 0.668 & 0.391 & \textbf{0.671} & \textbf{0.397} & 0.666 & 0.388 \\
2 & 2000 & 4  & 2 & \textbf{0.936} & \textbf{0.762} & 0.935 & 0.757 & \textbf{0.936} & \textbf{0.762} & \textbf{0.936} & 0.760 \\
3 & 3000 & 6  & 2 & 0.623 & 0.433 & \textbf{0.911} & \textbf{0.751} & 0.909 & 0.746 & \textbf{0.911} & 0.750 \\
4 & 4000 & 8  & 2 & 0.885 & 0.716 & \textbf{0.886} & \textbf{0.718} & 0.885 & 0.716 & 0.883 & 0.713 \\
5 & 5000 & 10 & 2 & \textbf{0.870} & \textbf{0.701} & 0.694 & 0.539 & 0.862 & 0.684 & 0.861 & 0.683 \\
6 & 6000 & 12 & 2 & \textbf{0.846} & \textbf{0.665} & 0.845 & 0.661 & 0.844 & 0.661 & 0.843 & 0.658 \\
7 & 7000 & 14 & 2 & 0.722 & 0.555 & 0.831 & 0.644 & \textbf{0.832} & \textbf{0.646} & 0.830 & 0.643 \\
8 & 8000 & 16 & 2 & \textbf{0.824} & \textbf{0.638} & 0.812 & 0.616 & 0.817 & 0.624 & 0.811 & 0.615 \\
9 & 9000 & 18 & 2 & \textbf{0.816} & \textbf{0.628} & 0.694 & 0.520 & 0.800 & 0.601 & 0.797 & 0.596 \\
\bottomrule
\end{tabular}
\end{table}

\begin{table}[H] 
\centering
\footnotesize
\begin{threeparttable}
\caption{Class 3: Gaussian data with simple Cauchy shifts. Accuracy and ARI for each dataset in this class. The highest Accuracy and ARI values in each row are shown in bold.}
\label{tab:simpc}
\begin{tabular}{cccccccccccc}
\toprule
  &      &    &    & \multicolumn{2}{c}{Kmeans} &
\multicolumn{2}{c}{Kmed} &
\multicolumn{2}{c}{KMR.L1} &
\multicolumn{2}{c}{KMR.L2}      \\
K & n    & p  & loc & Acc    & ARI   & Acc   & ARI   & Acc    & ARI   & Acc    & ARI    \\
2 & 2000 & 4  & 1.5 & 0.5 & 0 & 0.5 & 0 & 0.892 & 0.616 & 0.892 & 0.614 \\
3 & 3000 & 6  & 1.5 & 0.334 & 0 & 0.334 & 0 & \textbf{0.820} & \textbf{0.533} & 0.805 & 0.501 \\
4 & 4000 & 8  & 1.5 & 0.251 & 0 & 0.416 & 0.141 & \textbf{0.755} & \textbf{0.453} & 0.741 & 0.429 \\
5 & 5000 & 10 & 1.5 & 0.201 & 0 & 0.457 & 0.193 & \textbf{0.712} & \textbf{0.411} & 0.701 & 0.393 \\
6 & 6000 & 12 & 1.5 & 0.167 & 0 & 0.316 & 0.070 & 0.643 & 0.329 & \textbf{0.649} & \textbf{0.335} \\
7 & 7000 & 14 & 1.5 & 0.143 & 0 & 0.194 & 0.014 & 0.459 & 0.175 & \textbf{0.615} & \textbf{0.304} \\
8 & 8000 & 16 & 1.5 & 0.276 & 0.110 & 0.458 & 0.190 & 0.380 & 0.127 & \textbf{0.589} & \textbf{0.282} \\
9 & 9000 & 18 & 1.5 & 0.171 & 0.044 & 0.223 & 0.054 & 0.294 & 0.081 & \textbf{0.566} & \textbf{0.263} \\
2 & 2000 & 4  & 2 & 0.5 & 0 & 0.919 & 0.702 & 0.951 & 0.812 & 0.949 & 0.805 \\
3 & 3000 & 6  & 2 & 0.334 & 0 & 0.604 & 0.350 & \textbf{0.912} & \textbf{0.754} & 0.899 & 0.721 \\
4 & 4000 & 8  & 2 & 0.250 & 0 & 0.448 & 0.205 & \textbf{0.872} & \textbf{0.688} & 0.860 & 0.661 \\
5 & 5000 & 10 & 2 & 0.201 & 0 & 0.201 & 0 & \textbf{0.841} & \textbf{0.641} & 0.830 & 0.621 \\
6 & 6000 & 12 & 2 & 0.289 & 0.147 & 0.661 & 0.399 & \textbf{0.813} & \textbf{0.601} & 0.797 & 0.572 \\
7 & 7000 & 14 & 2 & 0.345 & 0.225 & 0.554 & 0.314 & \textbf{0.778} & \textbf{0.550} & 0.770 & 0.535 \\
8 & 8000 & 16 & 2 & 0.299 & 0.143 & 0.357 & 0.142 & \textbf{0.768} & \textbf{0.540} & 0.760 & 0.527 \\
9 & 9000 & 18 & 2 & 0.260 & 0.148 & 0.537 & 0.297 & 0.710 & 0.456 & \textbf{0.730} & \textbf{0.484} \\   \\
\end{tabular}

\begin{tabular}{cccccccccccc}
\toprule
  &      &    &      &
\multicolumn{2}{c}{Gauss} &
\multicolumn{2}{c}{t} &
\multicolumn{2}{c}{Cauchy} &
\multicolumn{2}{c}{G-Hpbl}       \\
K & n    & p  & loc  & Acc    & ARI   & Acc   & ARI   & Acc    & ARI   & Acc    & ARI     \\
2 & 2000 & 4  & 1.5  & 0.505      & 0          & 0.874    & 0.559    & 0.5        & 0          & \textbf{0.901}      & \textbf{0.643}      \\
3 & 3000 & 6  & 1.5  & 0.35       & 0.001      & 0.355    & 0.002    & 0.334      & 0          & 0.367      & 0.003      \\
4 & 4000 & 8  & 1.5  & 0.289      & 0.003      & 0.288    & 0.003    & 0.251      & 0          & 0.299      & 0.005      \\
5 & 5000 & 10 & 1.5  & 0.445      & 0.237      & 0.359    & 0.176    & 0.362      & 0.216      & DNC         & DNC         \\
6 & 6000 & 12 & 1.5  & 0.377      & 0.218      & 0.304    & 0.117    & 0.287      & 0.16       & DNC         & DNC         \\
7 & 7000 & 14 & 1.5  & 0.339      & 0.183      & 0.264    & 0.086    & 0.143      & 0          & DNC         & DNC         \\
8 & 8000 & 16 & 1.5  & 0.245      & 0.082      & 0.301    & 0.175    & 0.294      & 0.185      & 0.422      & 0.227      \\
9 & 9000 & 18 & 1.5  & 0.244      & 0.025      & 0.218    & 0.087    & 0.182      & 0.056      & 0.322      & 0.162      \\
2 & 2000 & 4  & 2    & 0.509      & 0          & 0.935    & 0.757    & \textbf{0.96}       & \textbf{0.844}      & 0.939      & 0.771      \\
3 & 3000 & 6  & 2    & 0.617      & 0.424      & 0.885    & 0.684    & 0.334      & 0          & DNC         & DNC         \\
4 & 4000 & 8  & 2    & 0.478      & 0.356      & 0.493    & 0.237    & 0.251      & 0          & DNC         & DNC         \\
5 & 5000 & 10 & 2    & 0.534      & 0.344      & 0.378    & 0.251    & 0.202      & 0          & 0.701      & 0.547      \\
6 & 6000 & 12 & 2    & 0.578      & 0.395      & 0.454    & 0.322    & 0.457      & 0.35       & 0.467      & 0.373      \\
7 & 7000 & 14 & 2    & 0.392      & 0.231      & 0.39     & 0.295    & 0.496      & 0.372      & 0.499      & 0.401      \\
8 & 8000 & 16 & 2    & 0.636      & 0.451      & 0.446    & 0.386    & 0.436      & 0.364      & DNC         & DNC         \\
9 & 9000 & 18 & 2    & 0.719      & 0.482       & 0.386    & 0.309    & 0.546      & 0.41       & DNC         & DNC    \\
\bottomrule
\end{tabular}
\begin{tablenotes}
\footnotesize
\item \textit{Note:} DNC denotes ``did not converge'' throughout the paper.
\end{tablenotes}
\end{threeparttable}
\end{table}

\begin{table}[hpt] 
\centering
\footnotesize
\setlength{\tabcolsep}{3pt}   
\renewcommand{\arraystretch}{0.95}
\caption{Class 4: Gaussian data with mixed Cauchy shifts, varying cluster sizes and densities. Accuracy and ARI for each dataset in this class. The highest Accuracy and ARI values in each row are shown in bold. Here, $K_1$ and $K_2$ denote the numbers of clusters in the two groups, $n_1$ and $n_2$ their respective total sample sizes, $pp_1$ and $pp_2$ the numbers of shifted coordinates for the clusters in groups $K_1$ and $K_2$, respectively, and $\mathrm{sep}_1$ and $\mathrm{sep}_2$ the corresponding separation parameters. Other notation follows Table~\ref{tab:class1}.}

\label{tab:2diff1}
\begin{tabular}{cccccccccccccccc}
\toprule
\multicolumn{2}{c}{K} &
\multicolumn{2}{c}{n} &
\multicolumn{2}{c}{pp} &
\multicolumn{2}{c}{sepp}  & \multicolumn{2}{c}{Kmeans} &
\multicolumn{2}{c}{Kmed} &
\multicolumn{2}{c}{KMR.L1} &
\multicolumn{2}{c}{KMR.L2}    \\
1 & 2 & 1 & 2    & 1 & 2  & 1  &  2 & Acc    & ARI   & Acc   & ARI   & Acc    & ARI   & Acc    & ARI     \\
2  & 2  & 2000 & 4000 & 3   & 3   & 1.5   & 1.5   & 0.573      & 0.219      & 0.657    & 0.353    & \textbf{0.818}      & \textbf{0.584}      & 0.804      & 0.559      \\
2  & 3  & 2000 & 6000 & 2   & 3   & 1.5   & 1.5   & 0.25       & 0          & 0.646    & 0.366    & \textbf{0.768}      & \textbf{0.530}       & 0.759      & 0.513      \\
2  & 4  & 2000 & 8000 & 2   & 3   & 1.5   & 1.5   & 0.2        & 0          & 0.47     & 0.192    & \textbf{0.742}      & \textbf{0.501}      & 0.733      & 0.483      \\
3  & 2  & 3000 & 4000 & 2   & 2   & 1.5   & 1.5   & 0.286      & 0          & 0.458    & 0.121    & \textbf{0.680}       & \textbf{0.370}       & \textbf{0.680}       & \textbf{0.370}       \\
3  & 3  & 3000 & 6000 & 3   & 2   & 1.5   & 1.5   & 0.223      & 0          & 0.341    & 0.082    & \textbf{0.690}       & \textbf{0.391}      & 0.674      & 0.369      \\
3  & 4  & 3000 & 8000 & 2   & 2   & 1.5   & 1.5   & 0.182      & 0          & 0.417    & 0.148    & 0.502      & 0.225      & \textbf{0.613}      & \textbf{0.311}      \\
4  & 2  & 4000 & 4000 & 2   & 2   & 1.5   & 1.5   & 0.251      & 0          & 0.414    & 0.132    & 0.624      & 0.316      & \textbf{0.626}      & \textbf{0.318}      \\
4  & 3  & 4000 & 6000 & 3   & 2   & 1.5   & 1.5   & 0.323      & 0.098      & 0.328    & 0.086    & \textbf{0.662}      & \textbf{0.363}      & 0.644      & 0.340      \\
4  & 4  & 4000 & 8000 & 3   & 2   & 1.5   & 1.5   & 0.167      & 0          & 0.278    & 0.059    & 0.536      & 0.231      & \textbf{0.615}      & \textbf{0.313}      \\
2  & 2  & 2000 & 4000 & 2   & 3   & 1.5   & 2     & 0.333      & 0          & 0.721    & 0.537    & \textbf{0.872}      & \textbf{0.725}      & 0.859      & 0.697      \\
2  & 3  & 2000 & 6000 & 2   & 2   & 1.5   & 2     & 0.25       & 0          & 0.635    & 0.353    & 0.771      & 0.546      & \textbf{0.785}      & \textbf{0.564}      \\
2  & 4  & 2000 & 8000 & 2   & 2   & 1.5   & 2     & 0.2        & 0          & 0.357    & 0.117    & 0.759      & 0.537      & \textbf{0.767}      & \textbf{0.544}      \\
3  & 2  & 3000 & 4000 & 2   & 2   & 1.5   & 2     & 0.286      & 0          & 0.517    & 0.216    & \textbf{0.750}       & \textbf{0.509}      & 0.748      & 0.504      \\
3  & 3  & 3000 & 6000 & 3   & 3   & 1.5   & 2     & 0.403      & 0.133      & 0.538    & 0.265    & \textbf{0.851}      & \textbf{0.702}      & 0.831      & 0.668      \\
3  & 4  & 3000 & 8000 & 2   & 2   & 1.5   & 2     & 0.183      & 0          & 0.582    & 0.327    & 0.626      & 0.406      & \textbf{0.717}      & \textbf{0.487}      \\
4  & 2  & 4000 & 4000 & 2   & 3   & 1.5   & 2     & 0.41       & 0.071      & 0.527    & 0.308    & \textbf{0.770}      & \textbf{0.589}      & 0.763      & 0.576      \\
4  & 3  & 4000 & 6000 & 2   & 3   & 1.5   & 2     & 0.346      & 0.075      & 0.435    & 0.215    & 0.761      & 0.594      & \textbf{0.769}      & \textbf{0.595}      \\
4  & 4  & 4000 & 8000 & 3   & 2   & 1.5   & 2     & 0.167      & 0          & 0.336    & 0.118    & \textbf{0.719}      & \textbf{0.480}      & 0.713      & 0.468     \\

\end{tabular}
\begin{tabular}{cccccccccccccccc}
\toprule
\multicolumn{2}{c}{K} &
\multicolumn{2}{c}{n} &
\multicolumn{2}{c}{pp} &
\multicolumn{2}{c}{sepp}  & \multicolumn{2}{c}{Gauss} &
\multicolumn{2}{c}{t} &
\multicolumn{2}{c}{Cauchy} &
\multicolumn{2}{c}{G-Hpbl}   \\
1 & 2 & 1 & 2    & 1 & 2  & 1  &  2 & Acc    & ARI   & Acc   & ARI   & Acc    & ARI   & Acc    & ARI     \\
2  & 2  & 2000 & 4000 & 3   & 3   & 1.5   & 1.5   & 0.474   & 0.297   & 0.448   & 0.216   & 0.626   & 0.359   & DNC     & DNC     \\
2  & 3  & 2000 & 6000 & 2   & 3   & 1.5   & 1.5   & 0.452   & 0.27    & 0.46    & 0.27    & 0.287   & -0.007  & 0.468   & 0.241   \\
2  & 4  & 2000 & 8000 & 2   & 3   & 1.5   & 1.5   & 0.572   & 0.412   & 0.38    & 0.22    & 0.202   & 0       & DNC     & DNC     \\
3  & 2  & 3000 & 4000 & 2   & 2   & 1.5   & 1.5   & 0.459   & 0.094   & 0.481   & 0.214   & 0.323   & -0.005  & DNC     & DNC     \\
3  & 3  & 3000 & 6000 & 3   & 2   & 1.5   & 1.5   & 0.427   & 0.202   & 0.334   & 0.131   & 0.223   & 0       & DNC     & DNC     \\
3  & 4  & 3000 & 8000 & 2   & 2   & 1.5   & 1.5   & 0.318   & 0.048   & 0.431   & 0.218   & 0.19    & 0.001   & 0.243   & 0       \\
4  & 2  & 4000 & 4000 & 2   & 2   & 1.5   & 1.5   & 0.325   & 0.062   & 0.427   & 0.157   & 0.251   & 0       & 0.334   & 0.159   \\
4  & 3  & 4000 & 6000 & 3   & 2   & 1.5   & 1.5   & 0.373   & 0.112   & 0.29    & 0.135   & 0.311   & 0.113   & 0.355   & 0.09    \\
4  & 4  & 4000 & 8000 & 3   & 2   & 1.5   & 1.5   & 0.292   & 0.028   & 0.397   & 0.207   & 0.175   & 0.001   & DNC     & DNC     \\
2  & 2  & 2000 & 4000 & 2   & 3   & 1.5   & 2     & 0.61    & 0.381   & 0.617   & 0.41    & 0.652   & 0.425   & 0.607   & 0.379   \\
2  & 3  & 2000 & 6000 & 2   & 2   & 1.5   & 2     & 0.675   & 0.477   & 0.647   & 0.441   & 0.469   & 0.314   & DNC     & DNC     \\
2  & 4  & 2000 & 8000 & 2   & 2   & 1.5   & 2     & 0.539   & 0.379   & 0.528   & 0.379   & 0.222   & -0.001  & DNC     & DNC     \\
3  & 2  & 3000 & 4000 & 2   & 2   & 1.5   & 2     & 0.402   & 0.265   & 0.516   & 0.284   & 0.286   & 0       & DNC     & DNC     \\
3  & 3  & 3000 & 6000 & 3   & 3   & 1.5   & 2     & 0.687   & 0.504   & 0.52    & 0.436   & 0.474   & 0.197   & 0.618   & 0.447   \\
3  & 4  & 3000 & 8000 & 2   & 2   & 1.5   & 2     & 0.642   & 0.454   & 0.353   & 0.194   & 0.341   & 0.205   & 0.622   & 0.435   \\
4  & 2  & 4000 & 4000 & 2   & 3   & 1.5   & 2     & 0.562   & 0.421   & 0.472   & 0.249   & 0.489   & 0.263   & DNC     & DNC     \\
4  & 3  & 4000 & 6000 & 2   & 3   & 1.5   & 2     & 0.634   & 0.51    & 0.548   & 0.34    & 0.576   & 0.399   & DNC     & DNC     \\
4  & 4  & 4000 & 8000 & 3   & 2   & 1.5   & 2     & 0.533   & 0.347   & 0.45    & 0.299   & 0.167   & 0       & DNC     & DNC     \\

\bottomrule
\end{tabular}
\end{table}

\newpage

\subsection{Clustering performance plots}

The following plots illustrate how clustering performance varies across the simulation settings, complementing the numerical results in the preceding tables.

\begin{figure}[H]
    \centering
\includegraphics[width=0.99\linewidth]{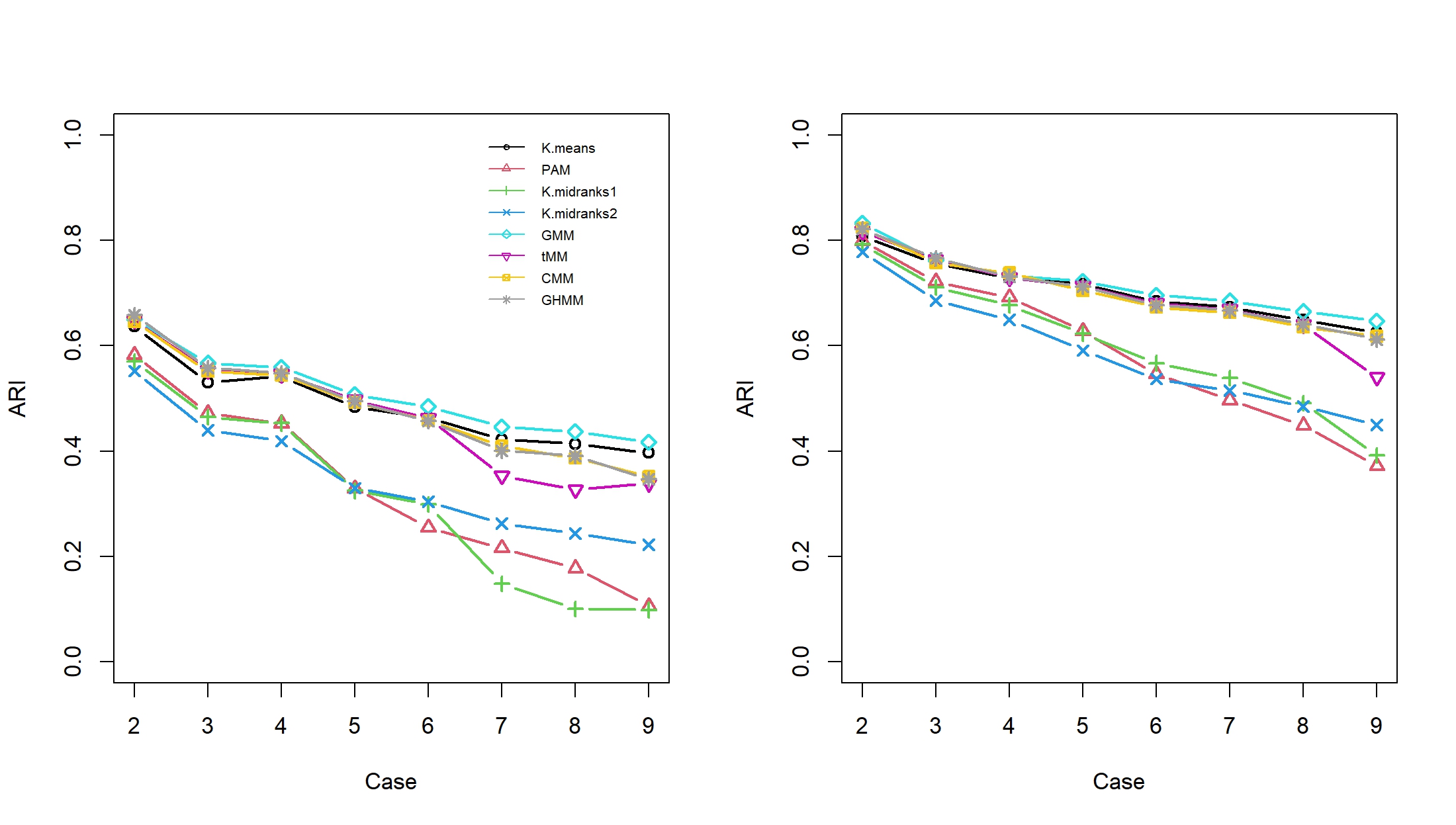}
    \caption{Class 1: Comparison of ARI values for $K=2$ to $K=9$ under separations of 1.5 (left) and 2 (right).}
    \label{fig:gau}
\end{figure}

\begin{figure}[H]
    \centering
\includegraphics[width=0.99\linewidth]{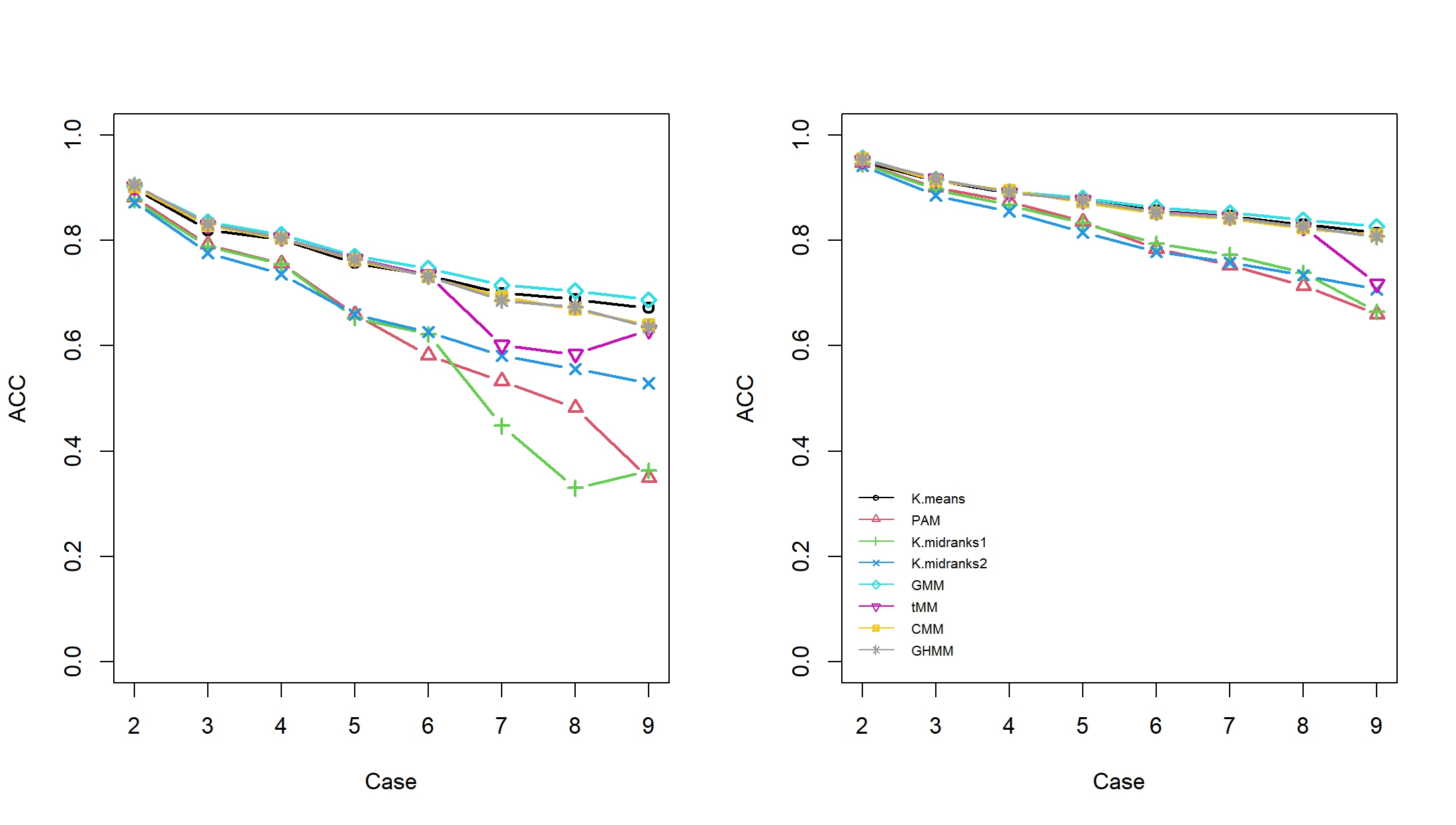}
    \caption{Class 1: Comparison of ACC values for $K=2$ to $K=9$ under separations of 1.5 (left) and 2 (right).}
    \label{fig:gau2}
\end{figure}

\begin{figure}[H]
    \centering
\includegraphics[width=0.99\linewidth]{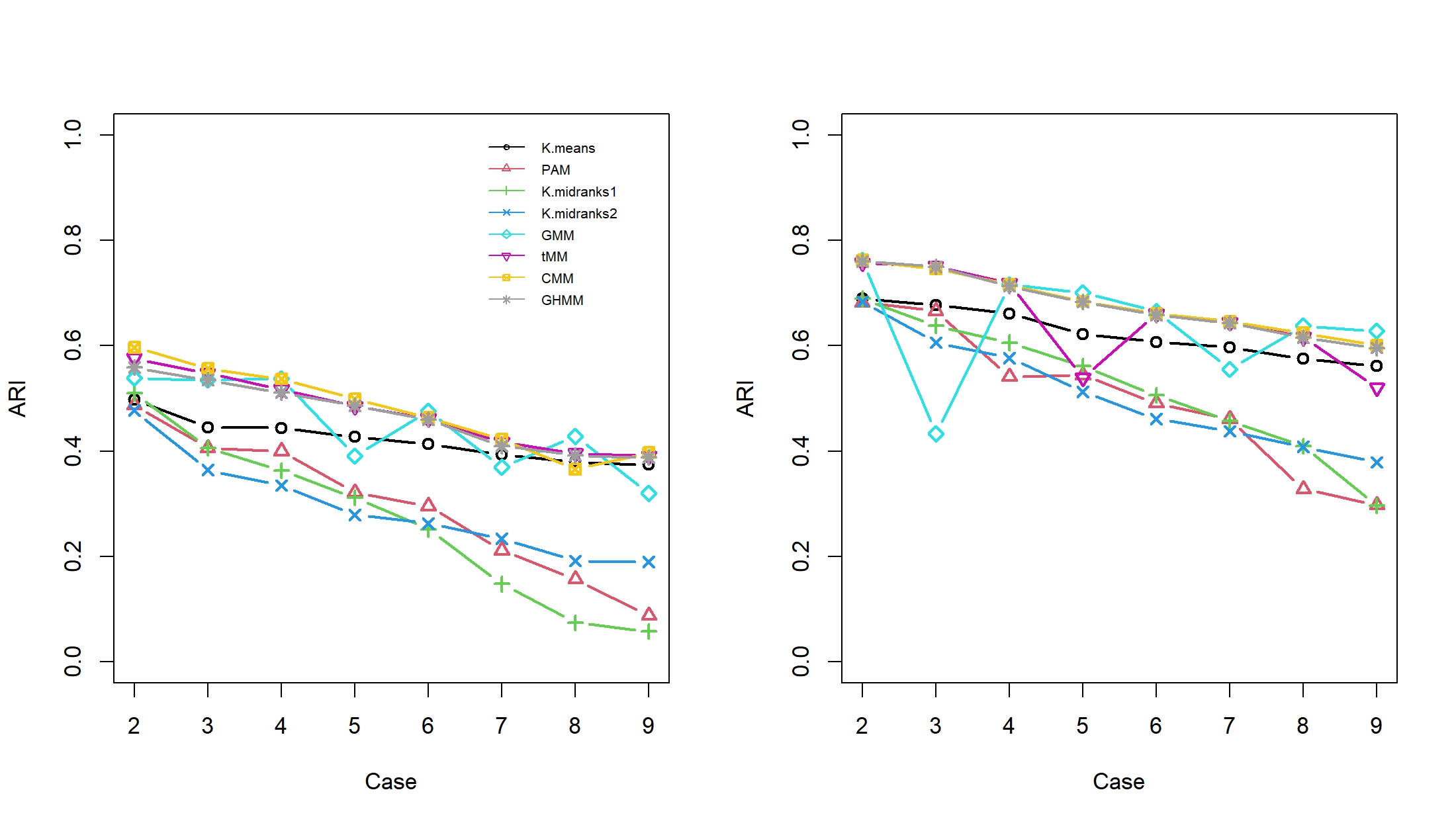}
    \caption{Class 2: Comparison of ARI values for $K=2$ to $K=9$ under separations of 1.5 (left) and 2 (right).}
    \label{fig:t}
\end{figure}

\begin{figure}[hbt]
    \centering
\includegraphics[width=0.99\linewidth]{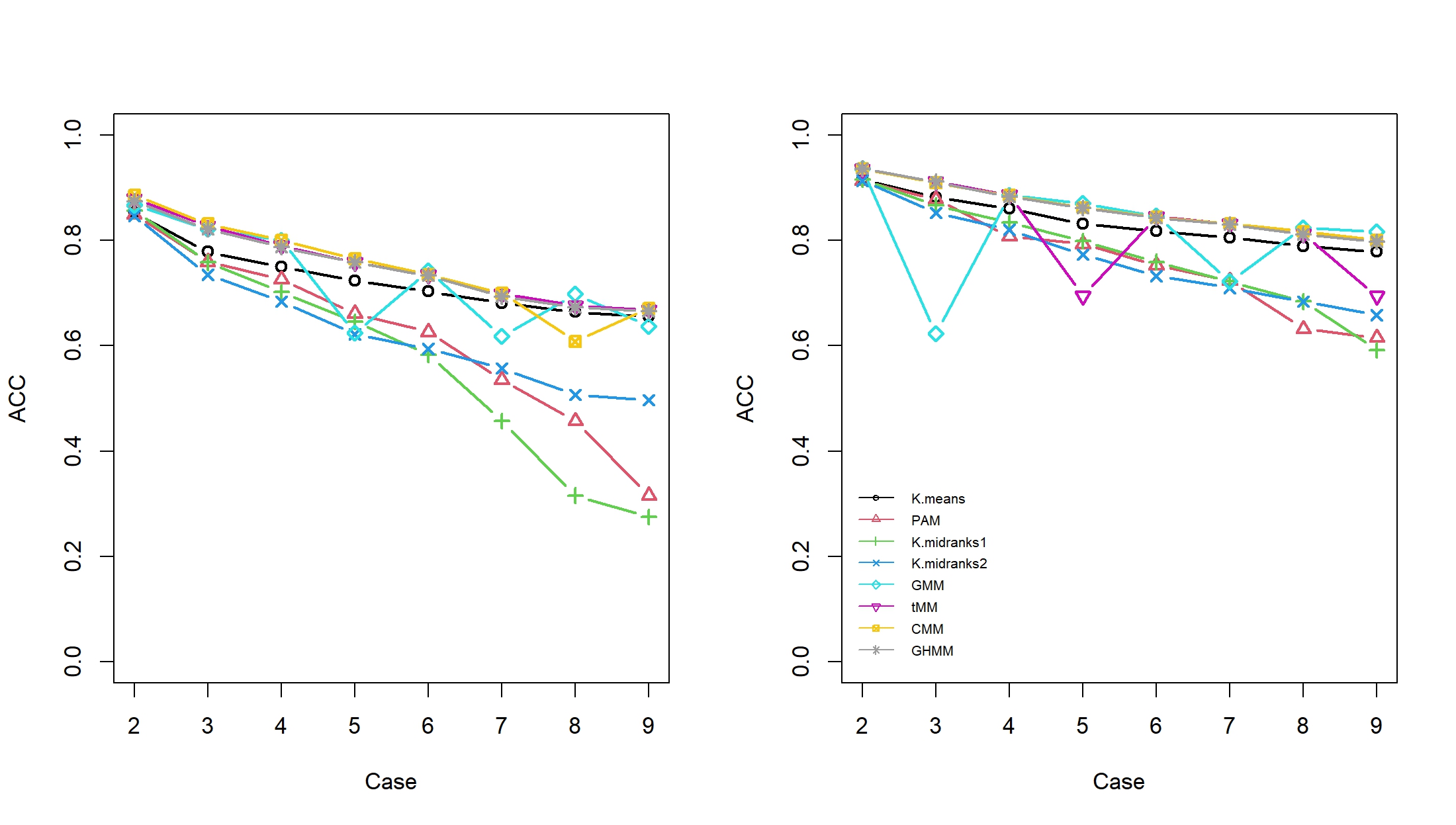}
    \caption{Class 2: Comparison of ACC values for $K=2$ to $K=9$ under separations of 1.5 (left) and 2 (right).}
    \label{fig:t2}
\end{figure}

\begin{figure}[H]
    \centering
\includegraphics[width=0.99\linewidth]{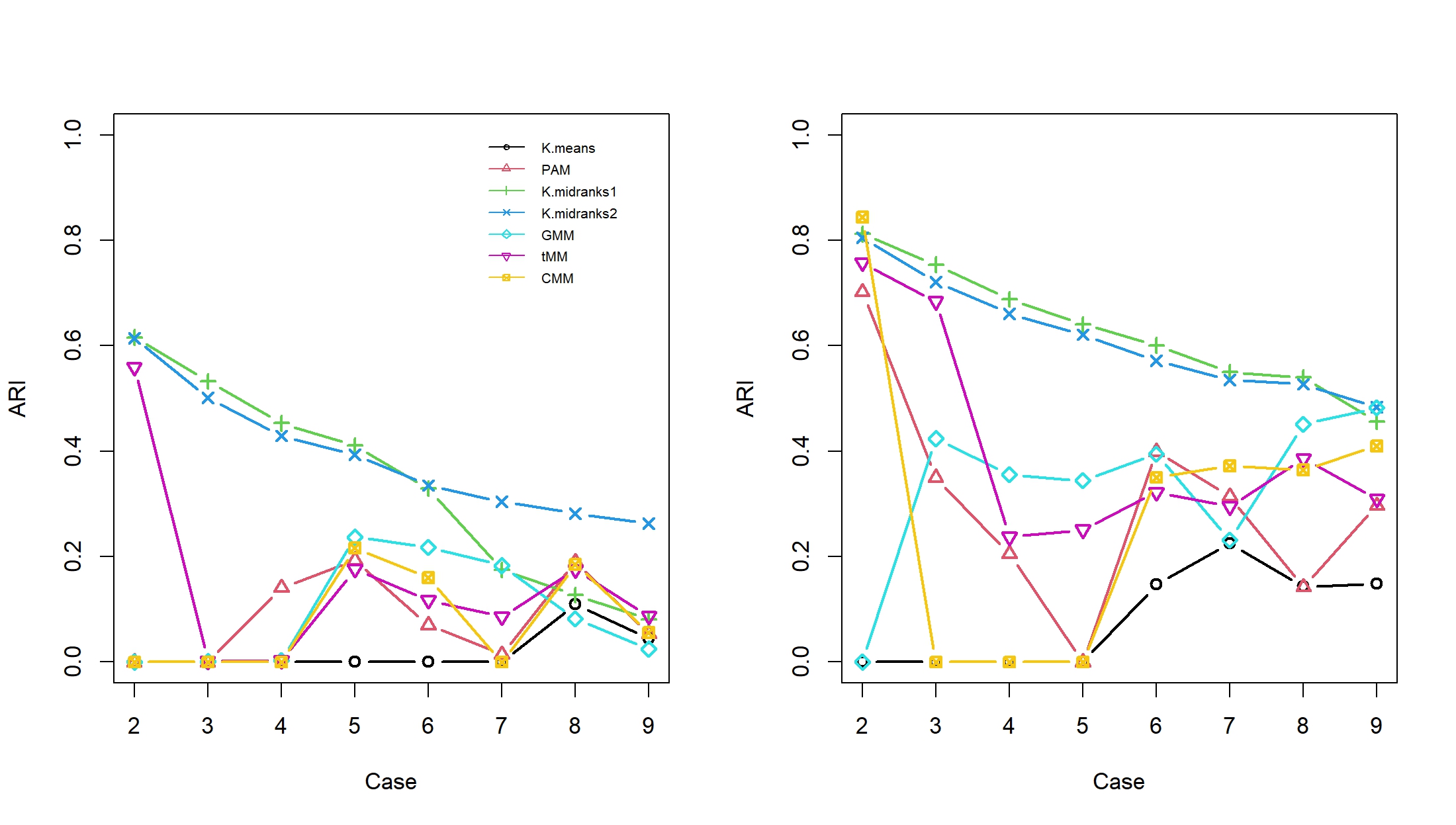}
   \caption{Class 3: Comparison of ARI values for $K=2$ to $K=9$ under separations of 1.5 (left) and 2 (right).}
    \label{fig:cau}
\end{figure}

\begin{figure}[hbt]
    \centering
\includegraphics[width=0.99\linewidth]{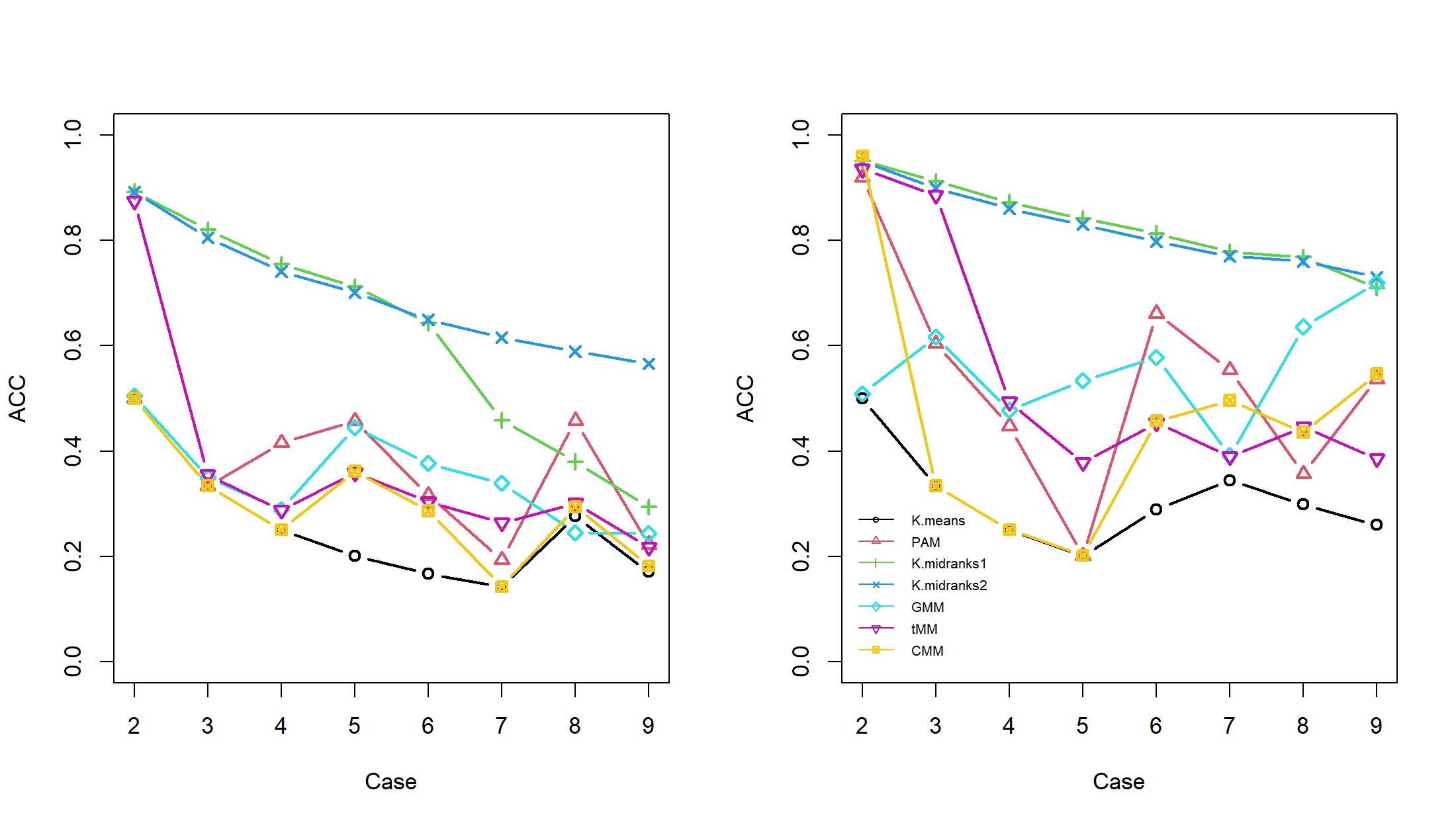}
   \caption{Class 3: Comparison of ACC values for $K=2$ to $K=9$ under separations of 1.5 (left) and 2 (right).}
    \label{fig:cau2}
\end{figure}

\begin{figure}[H]
    \centering
\includegraphics[width=0.99\linewidth]{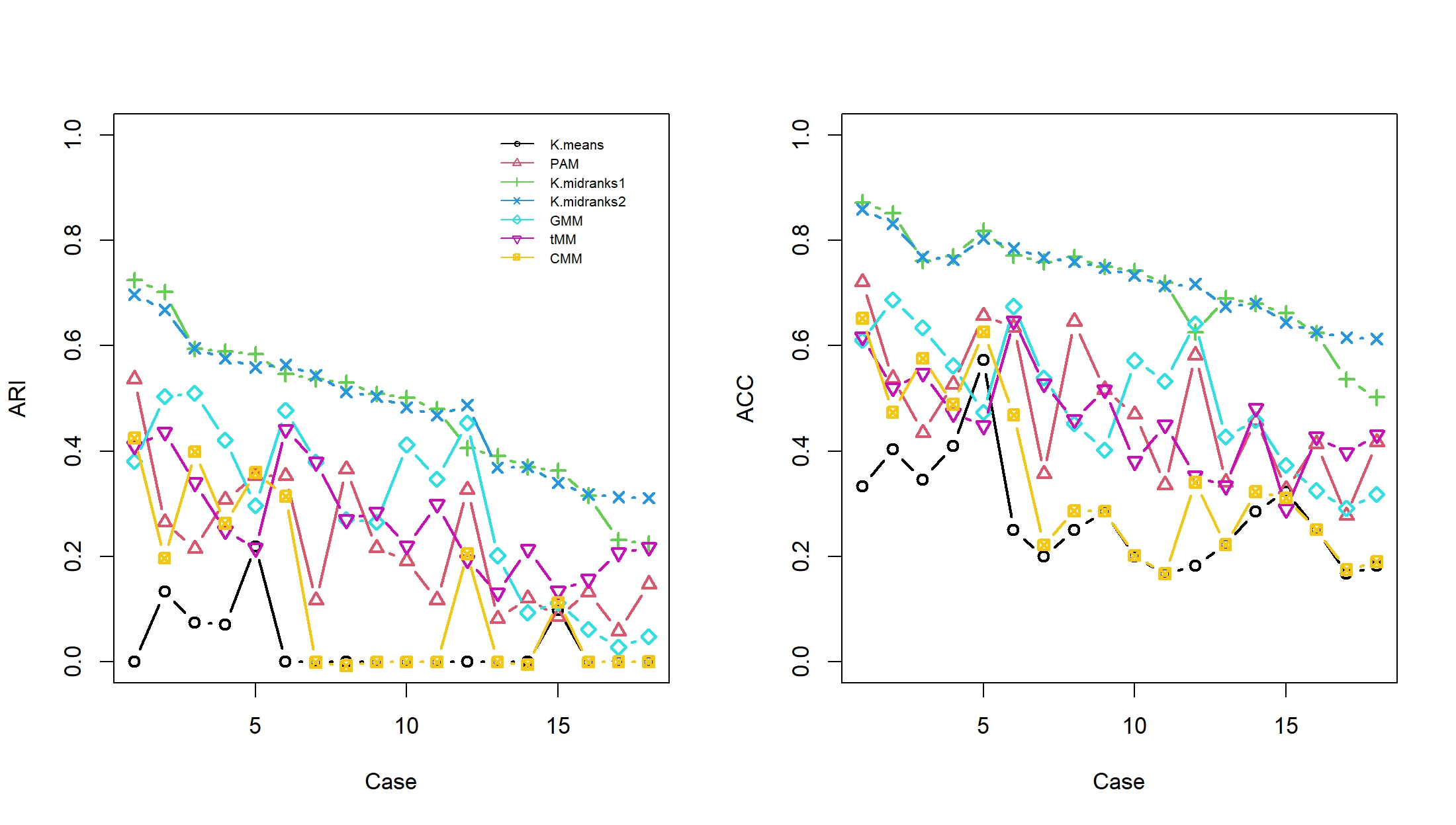}
    \caption{Class 4: ARI and ACC comparison ordered by the ARI values of $L^1$ $K$-midranks from largest to smallest.}
    \label{fig:class4}
\end{figure}

\subsection{Sensitivity analysis}

The following results examine the sensitivity of clustering performance to different random initializations.

\begin{table}[H] 
\centering
\footnotesize
\caption{Sensitivity analysis. Maximum, mean, standard deviation (SD), and minimum values of Accuracy and ARI over 30 runs for each dataset, used to assess the sensitivity of the algorithm to different initializations. Dist denotes the loss function, with 1 representing $L^1$ loss and 2 representing squared-Euclidean loss.}
\label{tab:sens}
\begin{tabular}{ccccccccccccc}
\toprule
  &    &  &    &      &
\multicolumn{4}{c}{Accuracy} &
\multicolumn{4}{c}{ARI}   \\
Loss & K & n    & p  & loc  & Max    & Mean   &  SD  & Min   & Max    & Mean   & sd    & Min     \\
1    & 2 & 2000 & 4  & 1.5  & 0.89    & 0.889    & 0.002  & 0.886   & 0.61    & 0.605    & 0.006  & 0.596   \\
1    & 2 & 2000 & 4  & 2    & 0.951   & 0.95     & <0.001      & 0.95    & 0.812   & 0.811    & 0.001  & 0.81    \\
1    & 5 & 5000 & 10 & 1.5  & 0.707   & 0.703    & 0.002  & 0.697   & 0.403   & 0.396    & 0.004  & 0.386   \\
1    & 5 & 5000 & 10 & 2    & 0.839   & 0.839    & <0.001      & 0.838   & 0.638   & 0.637    & 0.001  & 0.635   \\
1    & 9 & 9000 & 18 & 1.5  & 0.377   & 0.322    & 0.024  & 0.278   & 0.12    & 0.092    & 0.01   & 0.076   \\
1    & 9 & 9000 & 18 & 2    & 0.741   & 0.723    & 0.039  & 0.57    & 0.503   & 0.48     & 0.04   & 0.331   \\
2    & 2 & 2000 & 4  & 1.5  & 0.888   & 0.885    & 0.001  & 0.884   & 0.602   & 0.594    & 0.005  & 0.59    \\
2    & 2 & 2000 & 4  & 2    & 0.95    & 0.95     & <0.001      & 0.95    & 0.808   & 0.808    & <0.001      & 0.808   \\
2    & 5 & 5000 & 10 & 1.5  & 0.694   & 0.691    & 0.001  & 0.688   & 0.382   & 0.377    & 0.002  & 0.373   \\
2    & 5 & 5000 & 10 & 2    & 0.827   & 0.826    & <0.001      & 0.825   & 0.615   & 0.612    & 0.001  & 0.61    \\
2    & 9 & 9000 & 18 & 1.5  & 0.57    & 0.556    & 0.034  & 0.403   & 0.268   & 0.255    & 0.023  & 0.15    \\
2    & 9 & 9000 & 18 & 2    & 0.739   & 0.737    & 0.001  & 0.735   & 0.499   & 0.496    & 0.001  & 0.493  
\\
\bottomrule
\end{tabular}
\end{table}

\subsection{Estimation of the number of clusters}

The following table summarizes the performance of the procedures used to estimate the number of clusters in the simulated datasets.

\begin{table}[H] 
\centering
\footnotesize
\caption{Detecting the number of clusters. For each $K$, the table reports the $k$ values corresponding to the maximum and second maximum $B_q(k)$ values for $q=1,2$. The true numbers of clusters are shown in \textbf{bold}.}
\label{tab:elb1}
\begin{tabular}{cccccccc}
\toprule
n & p & loc & K & \shortstack{KMR.L1\\First choice} & \shortstack{KMR.L1\\Second choice} & \shortstack{KMR.L2\\First choice} & \shortstack{KMR.L2\\Second choice} \\
2000 & 4  & 1.5  & 2 & \textbf{2}     & 4        & \textbf{2}     & 4        \\
2000 & 4  & 2    & 2 & \textbf{2}     & 3        & \textbf{2}     & 3        \\
3000 & 6  & 1.5  & 3 & \textbf{3}     & 4        & \textbf{3}     & 8        \\
3000 & 6  & 2    & 3 & \textbf{3}     & 10       & \textbf{3}     & 6        \\
4000 & 8  & 1.5  & 4 & \textbf{4}     & 2        & \textbf{4}     & 7        \\
4000 & 8  & 2    & 4 & \textbf{4}     & 10       & \textbf{4}     & 8        \\
5000 & 10 & 1.5  & 5 & \textbf{5}     & 2        & \textbf{5}     & 2        \\
5000 & 10 & 2    & 5 & \textbf{5}     & 8        & \textbf{5}     & 2        \\
6000 & 12 & 1.5  & 6 & \textbf{6}     & 2        & \textbf{6}     & 2        \\
6000 & 12 & 2    & 6 & \textbf{6}     & 2        & \textbf{6}     & 2        \\
7000 & 14 & 1.5  & 7 & 10    & 2        & 10    & 2        \\
7000 & 14 & 2    & 7 & \textbf{7}     & 2        & \textbf{7}     & 2        \\
8000 & 16 & 1.5  & 8 & 2     & 4        & 2     & 3        \\
8000 & 16 & 2    & 8 & \textbf{8}     & 2        & \textbf{8}     & 2        \\
9000 & 18 & 1.5  & 9 & 2     & 4        & 2     & 4        \\
9000 & 18 & 2    & 9 & \textbf{9}     & 2        & 2     & \textbf{9}  
\\
\bottomrule
\end{tabular}
\end{table}

\subsection{Computational performance}

The following tables and plots summarize the computational performance of the methods, including runtimes and iteration counts across the simulation settings.

\begin{table}[H]
\centering
\footnotesize
\begin{threeparttable}
\caption{Computational results for the Class~3 datasets.
Subscripts 1 and 2 denote $K$-midranks with the $L^1$ and squared
Euclidean losses, respectively. For $j=1,2$, $\mathrm{rt}_j$ is the
observed runtime in seconds and $T_j$ is the number of iterations.
We compute $\mathrm{cp}_j=pn\log n+T_jnpK$ as given in the expression in
the run-time bound of Proposition~\ref{prop:kmr-current-algorithm}.}
\label{tab:computational-results}
\begin{tabular}{rrrrrrrrrrr}
\toprule
 & & & \multicolumn{2}{c}{Runtime (s)}
 & \multicolumn{2}{c}{Iterations}
 & \multicolumn{2}{c}{Complexity}
 & \multicolumn{2}{c}{Scaled runtime ratio} \\
$n$ & $p$ & $K$
 & $\mathrm{rt}_1$ & $\mathrm{rt}_2$
 & $T_1$ & $T_2$
 & $\mathrm{cp}_1$ & $\mathrm{cp}_2$
 & $10^9\mathrm{rt}_1/\mathrm{cp}_1$
 & $10^9\mathrm{rt}_2/\mathrm{cp}_2$ \\
\midrule
2000 &  4 & 2 & 0.04 & 0.02 &   6 & 11 &    156807.2 &   236807.2 & 255.090 & 84.457 \\
3000 &  6 & 3 & 0.03 & 0.02 &  20 & 10 &   1224114.6 &   684114.6 &  24.508 & 29.235 \\
4000 &  8 & 4 & 0.07 & 0.04 &  27 & 14 &   3721409.6 &  2057409.6 &  18.810 & 19.442 \\
5000 & 10 & 5 & 0.15 & 0.06 &  30 & 22 &   7925859.7 &  5925859.7 &  18.925 & 10.125 \\
6000 & 12 & 6 & 0.97 & 0.17 & 118 & 35 &  51602365.1 & 15746365.1 &  18.798 & 10.796 \\
7000 & 14 & 7 & 0.88 & 0.23 &  71 & 33 &  49573659.2 & 23505659.2 &  17.751 &  9.785 \\
8000 & 16 & 8 & 1.14 & 0.52 &  86 & 60 &  89214361.2 & 62590361.2 &  12.778 &  8.308 \\
9000 & 18 & 9 & 1.83 & 0.97 &  96 & 67 & 141443006.7 & 99161006.7 &  12.938 &  9.782 \\
\bottomrule
\end{tabular}
\end{threeparttable}
\end{table} 

\begin{figure}[H]
    \centering    \includegraphics[width=0.99\linewidth]{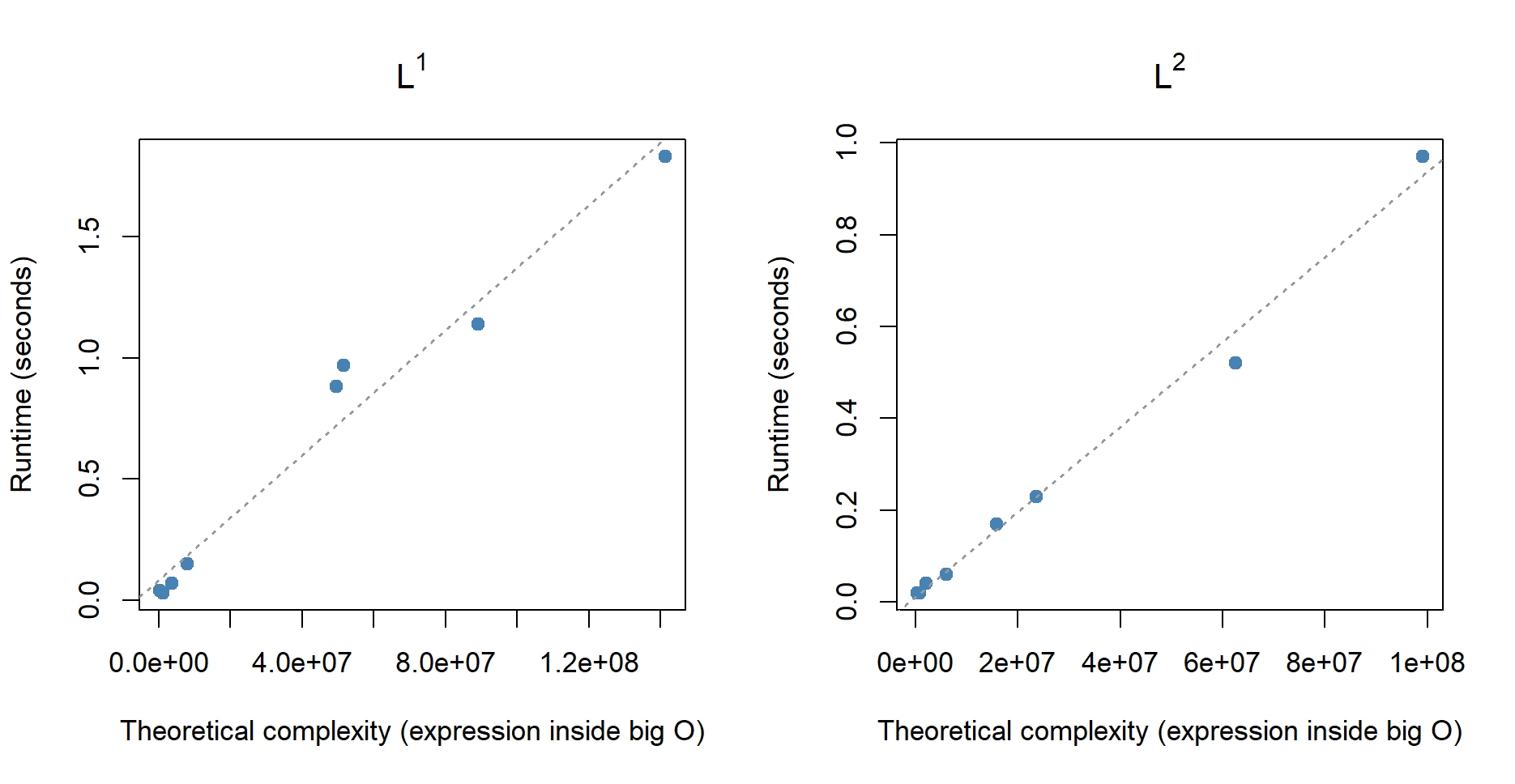}
    \caption{
Observed runtimes versus the expression $pn\log n+TnpK$ appearing in the theoretical computational complexity bound $O(pn\log n+TnpK)$ established in Proposition~\ref{prop:kmr-current-algorithm}, for the $L^1$ and $L^2$ $K$-midranks algorithms on the Class 3 datasets. Each point represents one dataset, and the dotted lines indicate fitted linear trends.
}
    \label{fig:runtime-complexity}
\end{figure}

\section{Real-data result}

This section contains the table summarizing clustering performance on the real datasets, and plots for the Raman spectroscopy laser-induced breakdown, and the S\&P 500 returns data. 

\subsection{Clustering performance}

The following table compares the clustering performance of the methods on the real datasets considered in this study.

\begin{table}[H] 
\centering
\footnotesize
\caption{Real-world data. Accuracy and ARI for each dataset. The highest Accuracy and ARI values in each row are shown in bold.}
\label{tab:rw}
\begin{tabular}{ccccccccccccc}
\toprule
  &      &    &  &  & \multicolumn{2}{c}{Kmeans} &
\multicolumn{2}{c}{Kmed} &
\multicolumn{2}{c}{KMR.L1} &
\multicolumn{2}{c}{KMR.L2}      \\
Data & n & p    & K  & PCA  & Acc    & ARI   & Acc   & ARI   & Acc    & ARI   & Acc    & ARI    \\
Raman & 3510   & 3276 & 32 & 20  & 0.499 & 0.354 & 0.569 & 0.407  & \textbf{0.829} & \textbf{0.733} & 0.759 & 0.698 \\
LIBs  & 10000  & 39995   & 12 & 50 & 0.215 & 0.060 & 0.217 & 0.060  & \textbf{0.536} & 0.303 & 0.526 & 0.306  \\
S\&P500 & 85 & 751 & 3 & NA  & 0.565 & 0.286 & 0.635 & 0.359  & 0.906 & 0.752 & \textbf{0.929} & \textbf{0.802}\\
S\&P500 & 215 & 751 & 5 & NA & 0.493 & 0.243 & 0.609 & 0.385 & 0.786 & 0.589 & \textbf{0.818} & \textbf{0.633}  \\
S\&P500 & 352 & 751 & 8 & NA  & 0.537 & 0.311 & 0.514 & 0.289 & \textbf{0.691} & \textbf{0.507} & 0.664 & 0.463   \\
\end{tabular}

\begin{tabular}{ccccccccccccc}
\toprule
  &  &    &       &  &
\multicolumn{2}{c}{Gauss} &
\multicolumn{2}{c}{t} &
\multicolumn{2}{c}{Cauchy} &
\multicolumn{2}{c}{G-Hpbl}       \\
Data & n & p    & K  & PCA  &   Acc    & ARI   & Acc   & ARI   & Acc    & ARI   & Acc    & ARI     \\
Raman & 3510 & 3276 & 32 & 20 & 0.789 & 0.656 & DNC & DNC & 0.681 & 0.664 & DNC & DNC \\
LIBs & 10000 & 39995 & 12 & 50 & 0.479 & 0.281 & 0.470 & 0.297 & 0.417 & 0.252 & 0.504 & \textbf{0.343} \\
S\&P500 & 85 & 751 & 3 & NA & 0.588 & 0.369 & DNC & DNC & DNC & DNC & DNC & DNC \\
S\&P500 & 215 & 751 & 5 & NA & 0.698 & 0.398 & DNC & DNC & DNC & DNC & DNC & DNC \\
S\&P500 & 352 & 751 & 8 & NA & 0.483 & 0.225 & DNC & DNC & DNC & DNC & DNC & DNC   \\
\bottomrule
\end{tabular}
\end{table}

\subsection{Raman spectroscopy}

The following plots illustrate the clustering results for the Raman spectroscopy data and their correspondence with the known substance labels.

\begin{figure}[H]
    \centering
\includegraphics[width=0.9\linewidth]{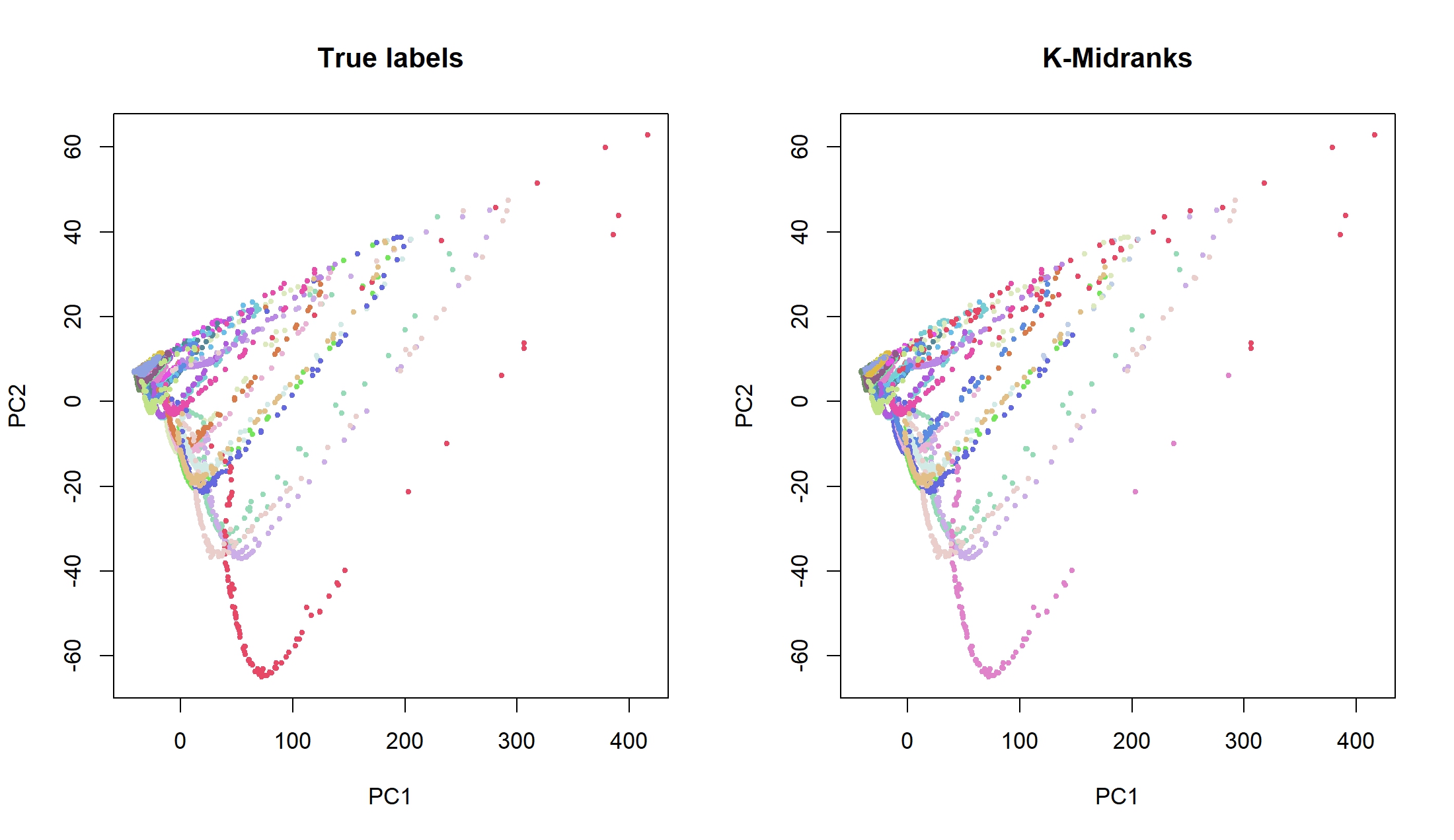}
    \caption{Projection of the true and $K$-midranks clustering labels onto the first two principal components of the Raman spectroscopy dataset. Each color represents one group.}
    \label{fig:raman}
\end{figure}

\begin{figure}[H]
    \centering
\includegraphics[width=0.99\linewidth]{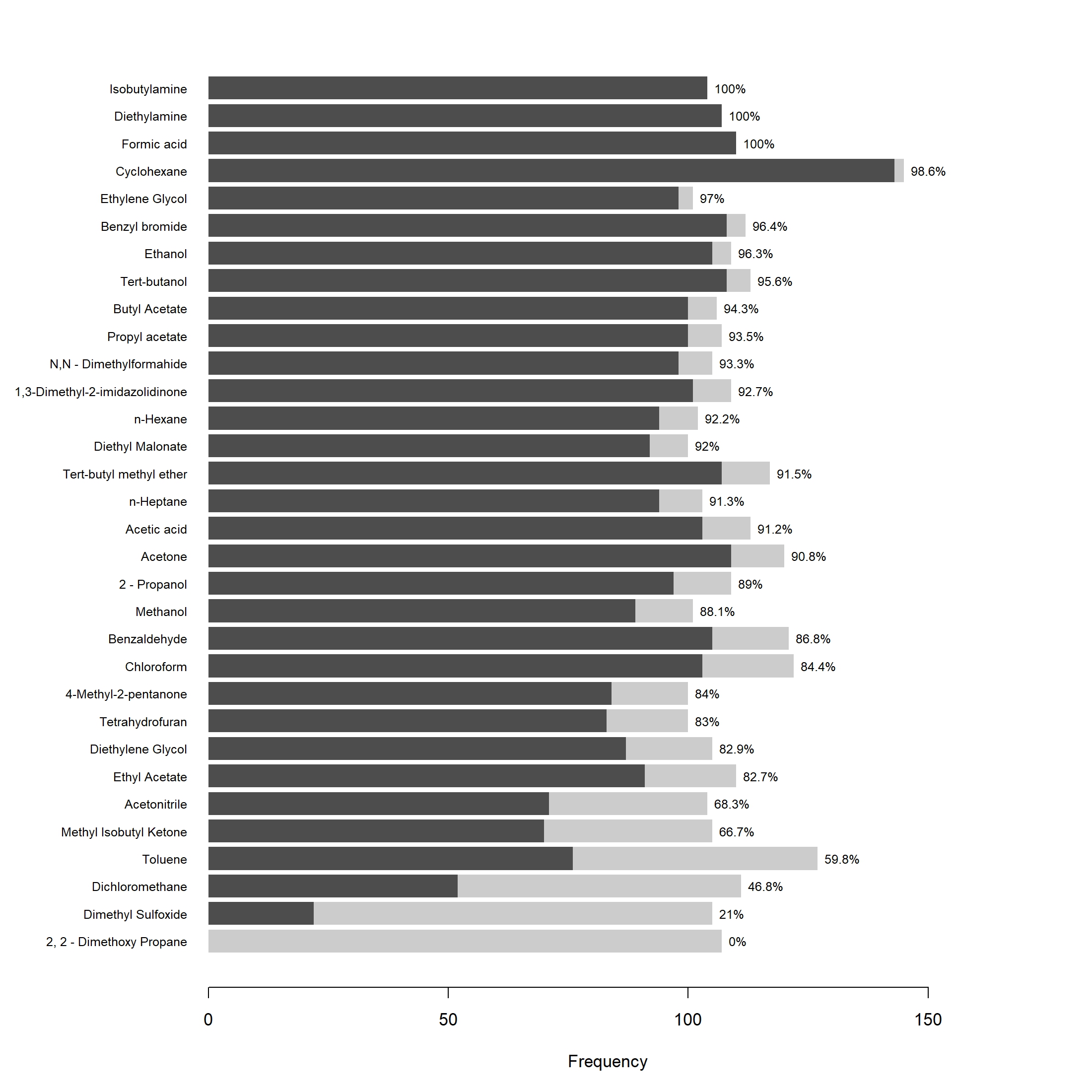}
    \caption{Frequency of true and false detections for each of the 32 materials in the Raman spectroscopy dataset using $K$-midranks with $L^1$ distance. Dark gray represents true detections, while light gray represents false detections. For each material, the detection percentage is calculated as the number of true detections divided by the total number of true and false detections for that material.}
    \label{fig:raman2}
\end{figure}

\subsection{Laser-induced breakdown spectroscopy}

Table~\ref{tab:libs-classes} lists the 12 geological classes and their corresponding numerical labels. The following plots illustrate the clustering results for the LIBS data and their correspondence with these known classes.

\begin{table}[ht!]
\centering
\caption{The 12 geological classes in the LIBS benchmark dataset of
\citet{kepevs2020benchmark} and their corresponding class labels.}
\label{tab:libs-classes}
\begin{tabular}{cl}
\toprule
\textbf{Class} & \textbf{Geological material} \\
\midrule
1  & U ore \\
2  & Zn-Pb-Ag sulfide ore \\
3  & Ni-Cu-PGE ore \\
4  & Sn ore \\
5  & Au-Cu ore \\
6  & Mn ore \\
7  & Gold oxide ore \\
8  & Zinc sulfide ore \\
9  & Hematite ore \\
10 & Anomalous ferruginous soil \\
11 & Silver-copper-gold ore \\
12 & Skarn tungsten magnetite ore \\
\bottomrule
\end{tabular}
\end{table}

\begin{figure}[H]
    \centering
\includegraphics[width=0.9\linewidth]{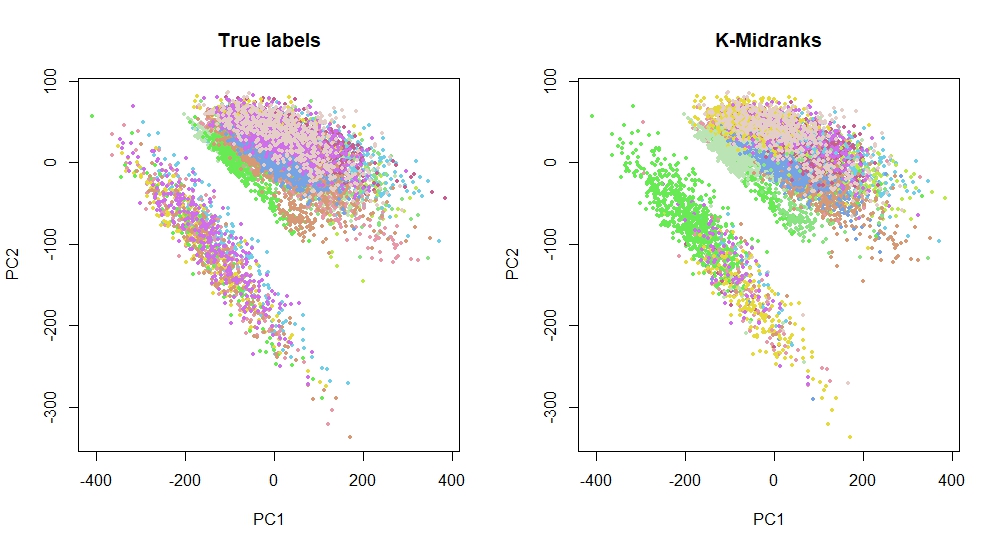}
    \caption{Projection of the true and $K$-midranks clustering labels onto the first two principal components of the LIBS dataset. Each color represents one cluster.}
    \label{fig:LIBs}
\end{figure}

\begin{figure}[H]
    \centering
\includegraphics[width=0.99\linewidth]{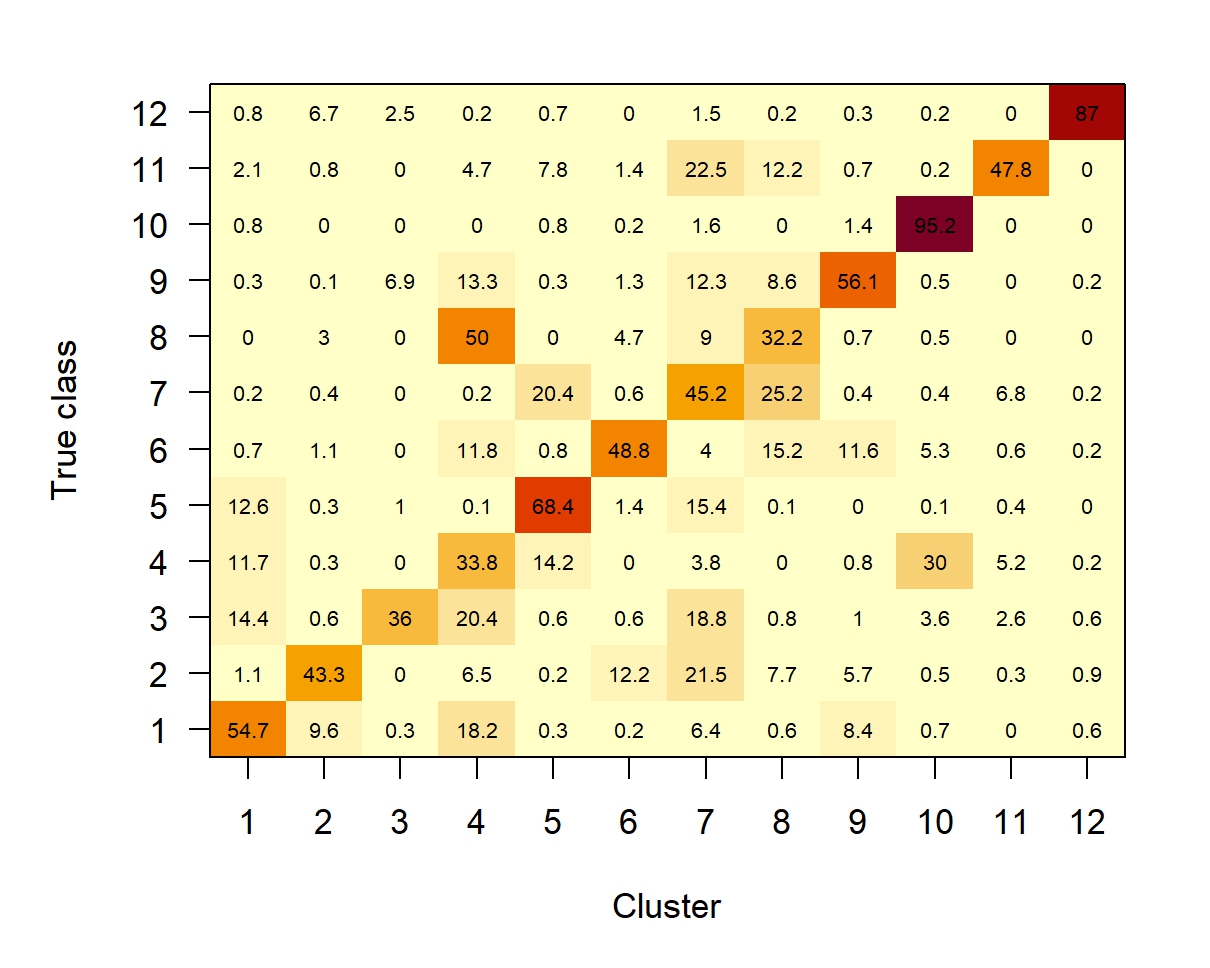}
    \caption{Heatmap of class-specific clustering results for the LIBS dataset using $K$-midranks with $L^1$ distance. Each row represents a true class and each column represents a matched cluster. The values indicate the percentage of observations from each true class assigned to each matched cluster, with each row summing to 100\%.}
    \label{fig:LIBs2}
\end{figure}

\subsection{S\&P 500 returns}

The following plots illustrate the clustering results for the S\&P 500 returns data and the corresponding closing-price trajectories.

\begin{figure}[H]
    \centering
\includegraphics[width=0.85\linewidth]{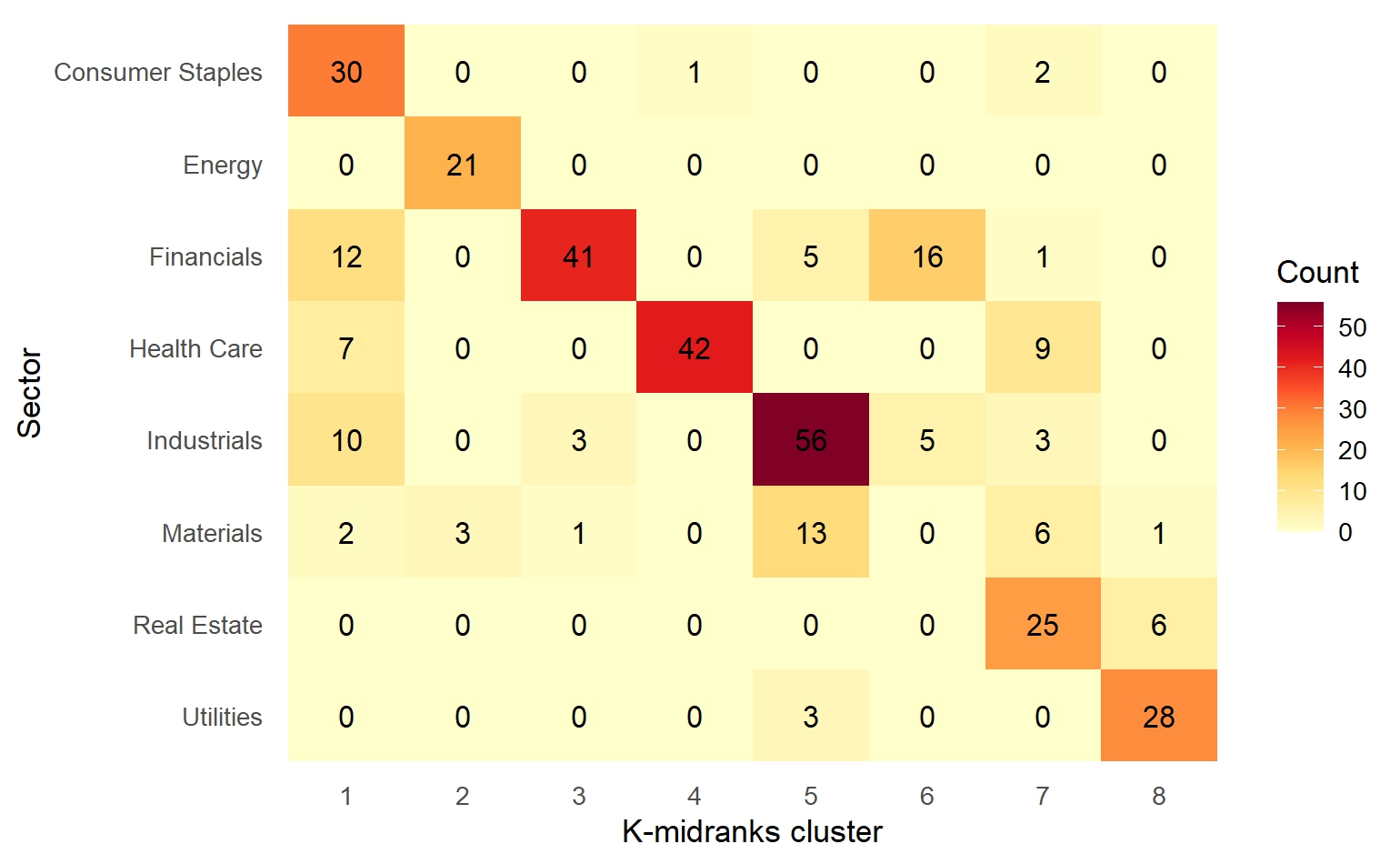}
    \caption{Correspondence between business sectors and clusters obtained using $L^1$ $K$-midranks in the eight-sector S\&P500 application. Rows represent sectors and columns represent clusters, with cell values indicating the number of companies. Cluster labels are reordered to maximize the sum of diagonal counts.}
    \label{fig:sphm}
\end{figure}  

\begin{figure}[H]
    \centering
\includegraphics[width=0.99\linewidth]{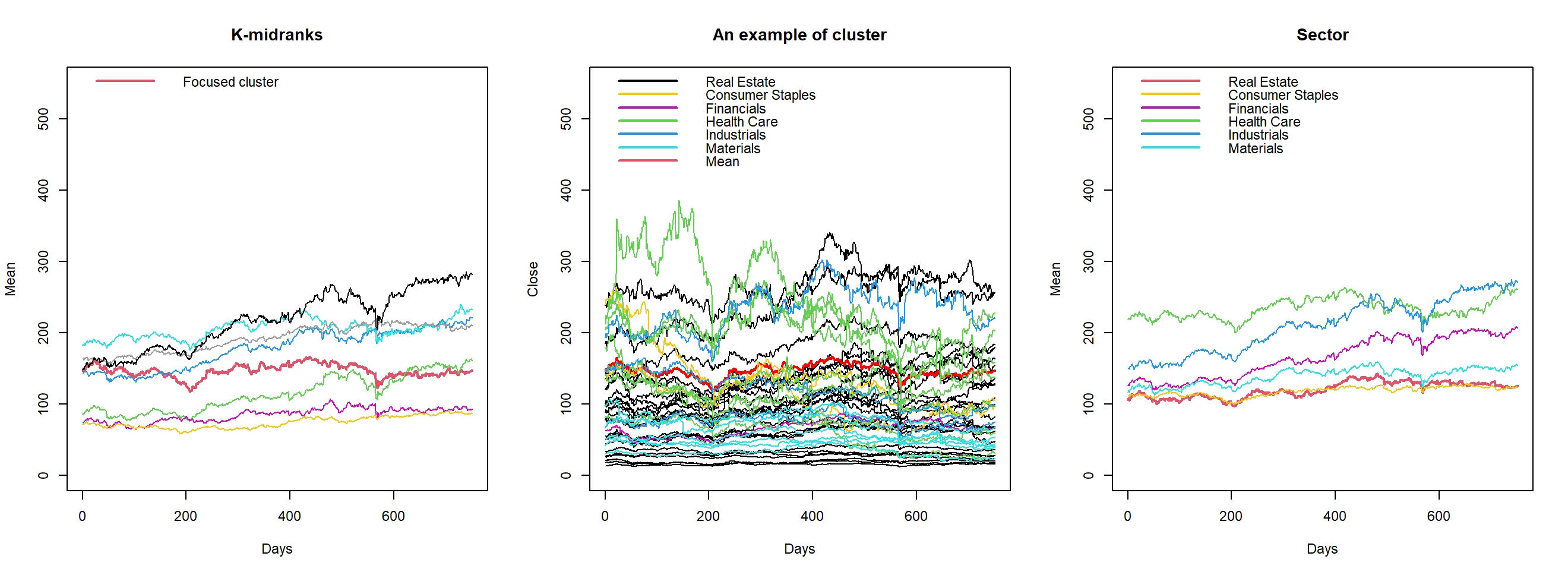}
    \caption{Illustration of $L^1$ $K$-midranks clustering in the eight-sector S\&P500 application. Left: mean daily closing-price trajectories of the eight clusters, with Cluster 7 highlighted in red. Middle: individual company trajectories in Cluster 7, colored by business sector, with the cluster mean in red. Companies with particularly high closing prices are omitted from the middle panel for visual clarity but are included in calculating the cluster mean. Right: sector-wide mean daily closing-price trajectories for the six sectors represented in Cluster 7.}
    \label{fig:sp}
\end{figure}  

\section{Normal Q–Q plots}
This section presents normal Q–Q plots for the simulated and real datasets to assess departures from normality, particularly differences in tail behavior. These plots provide distributional context for interpreting the clustering results discussed in the main text.

\begin{figure}[H]
    \centering
\includegraphics[width=0.9\linewidth]{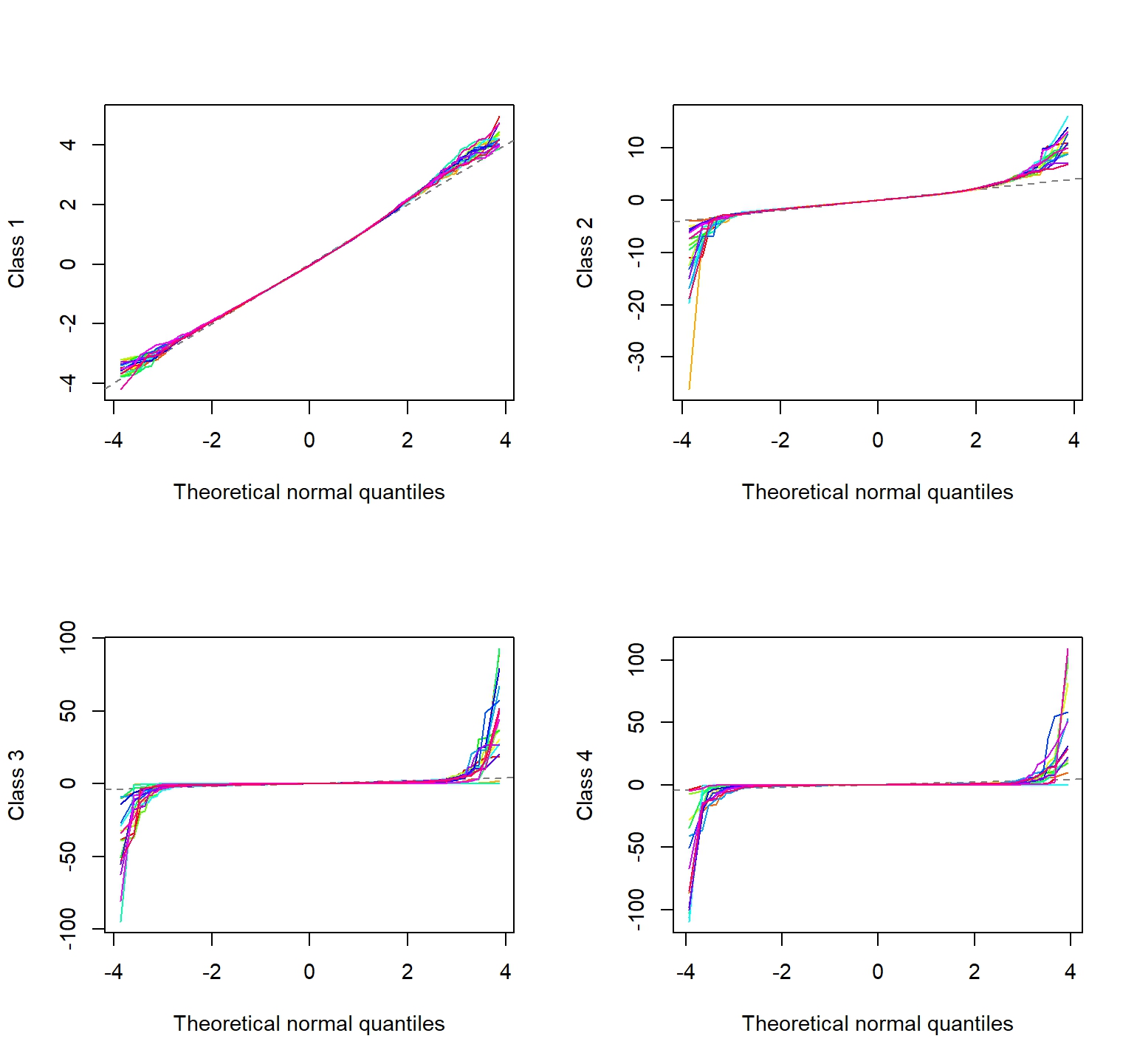}
    \caption{Normal Q--Q plots for all standardized variables from one selected case of each simulation class: Class 1 (Gaussian-based, $K=9$, $\mathrm{sepp}=1.5$), Class 2 ($t$-based, $K=9$, $\mathrm{sepp}=1.5$), Class 3 (Cauchy-based, $K=9$, $\mathrm{sepp}=1.5$), and Class 4 (Cauchy-based, $K_1=K_2=4$, $\mathrm{sepp}_1=\mathrm{sepp}_2=1.5$). Each colored curve is a Q--Q plot constructed from the observations of one variable across all clusters, and the dashed line represents the standard normal reference.}
    \label{fig:qq1}
\end{figure}

\begin{figure}[H]
    \centering
\includegraphics[width=0.99\linewidth]{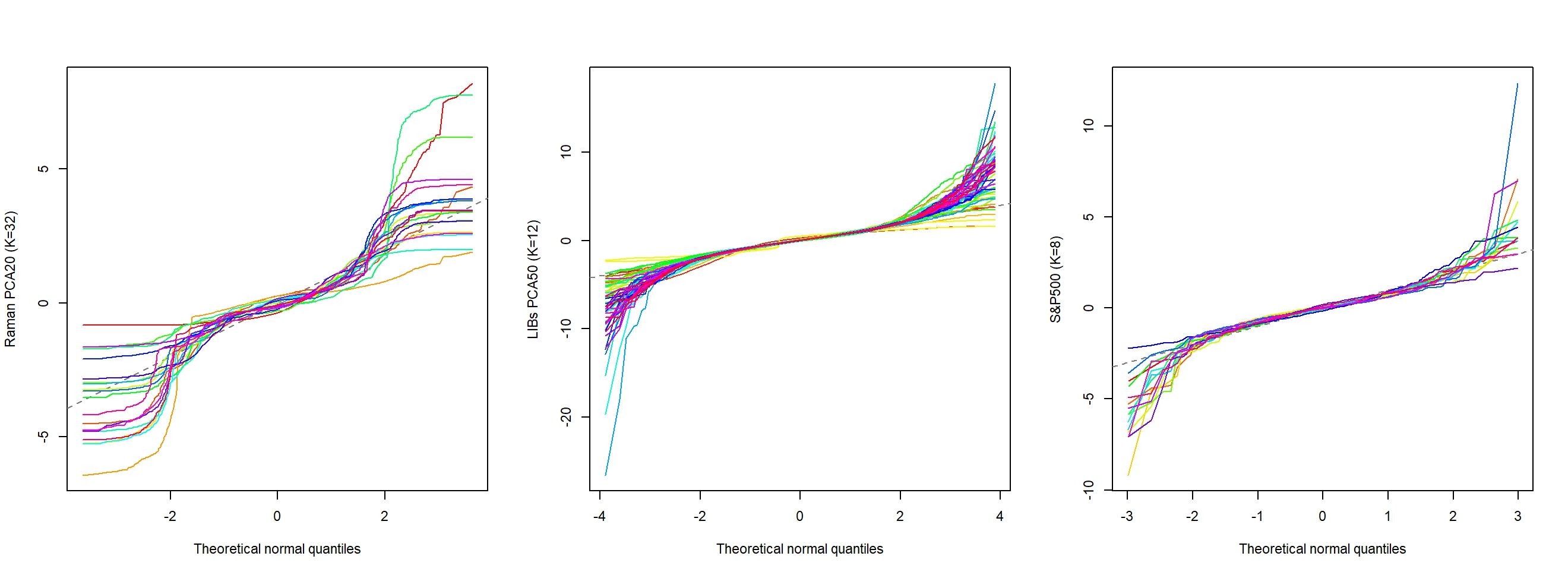}
    \caption{
Normal Q--Q plots for selected standardized variables from the real-world datasets: Raman with 20 retained principal components (left), LIBS with 50 retained principal components (center), and S\&P500 with 15 randomly selected time points (right). Each colored curve is a Q--Q plot constructed from the observations of one variable across all clusters, and the dashed line represents the standard normal reference.
}
    \label{fig:qq2}
\end{figure}



\bibliographystyle{imsart-nameyear}
\bibliography{ref}

\end{document}